\documentclass[conference,letterpaper]{IEEEtran}
\usepackage[top=0.73in,bottom=1.1in,left=0.637in,right=0.64in]{geometry}
\IEEEoverridecommandlockouts
\usepackage[utf8]{inputenc} 
\usepackage[T1]{fontenc}
\usepackage{url}
\usepackage{ifthen}
\usepackage{cite}
\usepackage[cmex10]{amsmath} % Use the [cmex10] option to ensure complicance
\usepackage{mleftright}       % fix to wrong spacing of \left-,
\mleftright                   % \middle- \right-commands 

\usepackage{graphicx}         % provides \includegraphics{...} to
\usepackage{booktabs}         % fixes poor spacing in tables and
\usepackage{dsfont}
\usepackage{svg}
\usepackage{upgreek}%ADDED by Prasad, Needed for some special greek fonts. 
\usepackage[linesnumbered]{algorithm2e}
\usepackage[shortlabels]{enumitem}
\usepackage{tikz}
\usepackage{lipsum,adjustbox}
\usetikzlibrary{calc,positioning}

\usepackage{amssymb}
\usepackage{amsfonts} 
\usepackage{tcolorbox}
\usepackage{enumitem}
\newlist{steps}{enumerate}{1}
\setlist[steps, 1]{label = Step \arabic*:}
\newcommand{\rl}{L}%Read length of Length of the read. Use this instead of L. 
\newcommand{\noofreads}{K} %No of reads. 
\newcommand{\bl}{n}%Block length of length of the entire sequence
\newcommand{\cover}{c}%Coverage, which is LK/n
\newcommand{\eras}{\perp}%Just defining this for erasure symbol. Can change if needed. Use this everywhere instead of \perp directly. 
 \newcommand{\ssechannel}{\mathsf{SSE}(\delta)}

\newcommand{\vx}{\underline{x}}%input vector
\newcommand{\vX}{\underline{X}}%random input vector
\newcommand{\vy}{\underline{y}}%Use  for reads.
\newcommand{\vynoerasures}{\tilde{\underline{y}}}%Use  for unerased reads.
\newcommand{\caly}{{\cal Y}}%to denote colection of reads with erasures(final output of channel)
\newcommand{\calynoerasures}{\tilde{\caly}}%to denote colection of reads after stage 1, without erasures. 
\newcommand{\expect}{\mathbb E}%Expectation
\newcommand{\indicatorRV}{\mathbb I}%indicator RV
\newcommand{\start}[1]{{\mathsf S}({#1})}
\newcommand{\startpos}[1]{{\cal S}_{#1}}

\newcommand{\B}{\boldsymbol} % Use bold when you are talking about random quantities. 

\newcommand{\normrl}{\bar{\rl}}%readlen/log\bl

\newcommand{\tth}{\texttt{th}}
\newcommand{\erasprob}{\delta}
\newcommand{\bigo}{\mathsf{O}}

\newcommand{\island}{Y'}
\newcommand{\ExLenIsland}{\Lambda}
\newcommand{\ekn}{ } %for $e_{k,\bl}$
\newcommand{\recalc}[1]{{{#1}}}
\newtheorem{lemma}{Lemma}

\newtheorem{claim}{Claim}
\newtheorem{definition}{Definition}
\newtheorem{theorem}{Theorem}

\RestyleAlgo{ruled}

\title{A Converse For the Capacity of the Shotgun Sequencing Channel with Erasures}
\author{Mohammed Ihsan Ali, Hrishi Narayanan, Prasad Krishnan %
\thanks{Ali and Krishnan are with the Signal Processing and Communications Research Center, International Institute of Information Technology, Hyderabad 500032, India (emails: \{mohammed.ihsan@students., prasad.krishnan@\}iiit.ac.in). Narayanan is at the Institute for Communications Engineering, Technical University of Munich, Germany (email: hrishi.narayanan@tum.de).  
}%
\thanks{Krishnan's contribution to this work was supported by ANRF via grant CRG/2023/08696.}

}

\begin{document}

\maketitle

%%% Single author, or several authors with same affiliation:
% \author{%
 % \IEEEauthorblockN{Hrishi Narayanan, Prasad Krishnan, Nita Parekh}
 % \IEEEauthorblockA{\\IIIT Hyderabad\\
%                    Emails: \{hrishi.narayanan@research., prasad.krishnan@, nita@\}iiit.ac.in}
% }

%%% Several authors with up to three affiliations:
% \author{%
%   \IEEEauthorblockN{Hrishi Narayanan}
%   \IEEEauthorblockA{IIIT Hyderabad\\
%                     Hyderabad, Telangana, India\\
%                     Email: hrishi.narayanan@research.iiit.ac.in}
%   \and
%   \IEEEauthorblockN{Prasad Krishnan}
%   \IEEEauthorblockA{Signal Processing and Communication Research Centre\\
%                     IIIT Hyderabad\\
%                     Hyderabad, Telangana, India\\
%                     Email:  prasad.krishnan@iiit.ac.in}
%   \and
%   \IEEEauthorblockN{Nita Parekh}
%   \IEEEauthorblockA{Centre for Computational Natural Sciences and Bioinformatics\\
%                     IIIT Hyderabad\\
%                     Hyderabad, Telangana, India\\
%                     Email:  nita.parekh@iiit.ac.in}
% }

%%%%%%
%% Abstract: 
%% If your paper is eligible for the student paper award, please add
%% the comment "THIS PAPER IS ELIGIBLE FOR THE STUDENT PAPER
%% AWARD." as a first line in the abstract. 
%% For the final version of the accepted paper, please do not forget
%% to remove this comment!
%%

\begin{abstract} 
    The shotgun sequencing process involves fragmenting a long DNA sequence (input string) into numerous shorter, unordered, and overlapping segments (referred to as \emph{reads}). The reads are sequenced, and later aligned to reconstruct the original string. Viewing the sequencing process as the read-phase of a DNA storage system, the information-theoretic capacity of noise-free shotgun sequencing has been characterized in literature. Motivated by the base-wise quality scores available in practical sequencers, a recent work considered the \emph{shotgun sequencing channel with erasures}, in which the symbols in the reads are assumed to contain random erasures. Achievable rates for this channel were identified. In the present work, we obtain a converse for this channel. The arguments for the proof involve a careful analysis  of a genie-aided decoder, which knows the correct locations of the reads. The converse is not tight in general. However, it meets the achievability result asymptotically in some channel parameters.
\end{abstract}

\section{Introduction}
In recent years, considerable academic as well as industrial interest has been observed in using artificially synthesized DNA molecules as a medium for data storage. In particular, due to their higher data stability and density, DNA-based storage systems show great promise as a medium for archival storage \cite{carmean_hybrid_moelcular_electronic_computing,takahashi_end_to_end,heckel_characterization_DNA_storage}. The fundamental techniques underlying DNA-based data storage systems are DNA synthesis and sequencing. A major constraint in DNA sequencing is the restriction on the maximum read length that the sequencer is capable of sequencing. An important development that addressed this was the invention of the \emph{high-throughput shotgun sequencing} pipeline. In this process, rather than sequencing the DNA sequence as is, multiple copies of the sequence are made and broken using restriction enzymes to obtain overlapping fragments. These fragments are much smaller than the original sequence in length and are subsequently `read' by the sequencer, resulting in the collection of \emph{reads}. The original sequence is then algorithmically reconstructed by aligning the reads by mapping their overlaps. This is a notoriously difficult problem, as the reads are numerous but their ordering (i.e., locations in the original string where they appear) is unknown. Several heuristic and analytical approaches exist that address this problem and its fundamental limits (see, for instance, \cite{bresler_optimal_assembly, shomorony_info_optimal_genome_assembly}). 

An early work which mathematically modeled and analyzed DNA sequence assembly was by Lander and Waterman \cite{lander_genomic_mapping}, established several limits on the sequencing parameters necessary for reliable reconstruction. Building on this work, the problem of shotgun sequencing was approached from an information theory perspective by Motahari \emph{et al.} \cite{motahari_info_theory_dna-shotgun_sequencing}. 
The work demonstrated some necessary and sufficient conditions on the parameters of the sequencer for the reconstruction of the input sequence, in the asymptotic regime (in the length of the input). The framework and approach in \cite{motahari_info_theory_dna-shotgun_sequencing} were used in later works studying the fundamental limits of DNA storage systems, with the goal of establishing the \emph{capacity} of the DNA sequencing channel. For instance, \cite{heckel_fundamental_limits, Shomo_Heckel_TIT_ShufflingSampling_2021, shomorony_torn_paper, ravi_recovering_message_incomplete_set} analyze the capacity in the regime where the reads are non-overlapping. The capacity of the \emph{shotgun sequencing channel} was presented in \cite{ravi_coded_ssc}. In the channel model considered in \cite{ravi_coded_ssc}, $\noofreads$ reads of fixed length $\rl = \normrl \log{\bl}$ are obtained by uniformly sampling starting positions from the indices of the input string $\vx$. The capacity of this channel was shown to be $(1-e^{-c(1-1/{\normrl})})$, where $c=\frac{\noofreads\rl}{\bl}$ denotes the coverage depth (which captures the average number of times any position in $\vx$ occurs in the collection of reads). However, this work considered the reads to be free of any errors. A related series of works \cite{ElischoGabrysYaakobi_TIT_RepeatFree,MarkovichYaakobi_TIT_ReconstructSubstringSpect} considers the string reconstruction problem from its \textit{$L$-multispectrum}. The \textit{$L$-multispectrum} of a string is the multiset of all possible $L$-length (contiguous) substrings of the input string, taken from every possible starting position. These works provide results on the capacity of specific multi-spectrum channels which are, however, distinct from the model in \cite{ravi_coded_ssc}. Another active line of work along these lines is in the area of information-theoretic analysis of DNA storage using nanopore sequencing. A number of works, including some very recent ones, have been done in this model (for instance, \cite{weidiggavisreeramTIT2018nanopore,reyna2021nanoporeISIT,mcbainviterbonanopore2024TIT}). 

A recent work \cite{hrishietal_SEEachieva_ISIT_2024,arxiv_Hri_SSE_24}, analyzed the shotgun sequencing channel under erasure errors $\ssechannel$. This was motivated by the fact that many modern sequencers provide quality scores (indicating the probability that the base has been correctly identified) corresponding to each base in the sequence, and by appropriately thresholding the quality score, one can consider bases with low quality scores to be erased. The work \cite{hrishietal_SEEachieva_ISIT_2024} presented an achievability result for $\ssechannel$, thereby effectively providing a lower bound for the channel capacity.

Random coding arguments are used to show the achievability results in \cite{ravi_coded_ssc,hrishietal_SEEachieva_ISIT_2024}. A tight converse is proved for the error-free shotgun sequencing channel in \cite{ravi_coded_ssc}. A main ingredient in the proof of the converse is the notion of `islands', which refers to contiguous maximal substrings of the original string that can be reconstructed by aligning the reads in appropriate order. The information about the original input sequence that is contained in the reads, thus, is completely captured in the islands as well. These islands are non-overlapping, as they emerge subsequent to the merging of reads. Post these observations, the work \cite{ravi_coded_ssc} adopts the framework of a \textit{torn-paper channel} (TPC) \cite{shomorony_torn_paper,ravi_torn_paper_lost_pieces} to complete the converse. The TPC is another channel model, wherein the reads obtained from the input sequence are themselves non-overlapping. The capacity of the TPC has been characterized \cite{shomorony_torn_paper,ravi_torn_paper_lost_pieces}, wherein it was observed that the capacity expression is a difference between two terms, (i) a `coverage' term which gives the expected number of bits of the input that appear in the reads, and (ii) a `re-ordering cost', which gives the number of bits needed for reordering the reads.
% the cost paid in bits, to reorder the reads. 
By establishing a rigorous analogy between the islands of the noise-free shotgun channel and the non-overlapping reads of the TPC, the authors of \cite{ravi_coded_ssc} obtain a tight converse for the shotgun channel. 

In the context of the shotgun channel with erasures, the work \cite{hrishietal_SEEachieva_ISIT_2024} provides an achievability result. While it is conceivable that a converse may be obtained using arguments along the lines of \cite{ravi_coded_ssc} (via results for an erasure-prone TPC), so far such an attempt seems absent in literature. A longer version of the work \cite{ravi_torn_paper_lost_pieces}, the authors present achievability and converse results for a noisy TPC, where the noise considered is independent bit-flip errors on the non-overlapping reads. However, for obtaining a converse for the shotgun channel with erasures, results on the erasure-noisy torn-paper channel are needed. Further, in such a model, if erasures on different bits of the TPC-reads are modeled as independent, this would also not suffice for our purpose, since the analogous erasures seen in the shotgun-islands are \textit{not} independent (though the erasures in the shotgun-reads are). Thus, addressing the issue of finding converses for the erasure-prone shotgun channel remains open. 

This work presents a converse for the shotgun sequencing channel with erasures. The arguments adopted to establish this converse are along the lines of those in \cite{ravi_coded_ssc} and \cite{ravi_recovering_message_incomplete_set}. However, the arguments are not trivial extensions of the same, as the analysis is more involved in several places. Further, while the arguments in \cite{ravi_coded_ssc} for the noise-free case relies on those for the TPC \cite{shomorony_torn_paper,ravi_torn_paper_lost_pieces}, this connection is not very transparently established. In this work, we present the entire proof for the converse for the shotgun channel with erasures, under one roof, in a streamlined fashion (to the best of our ability). In the process, we also clearly bring out the choices of the parameters involved in the analysis, and the structure of the complete proof, which in our opinion, seems difficult to see in \cite{ravi_coded_ssc}. The converse we obtain is not tight in general. However, the achievability result from \cite{hrishietal_SEEachieva_ISIT_2024} and the converse result here both approach $1$ as the coverage depth $c$ goes to $\infty$. 

This paper is organised as follows. In Section \ref{section:Channel description}, we restate the channel model and recall some preliminary definitions and results from \cite{ravi_coded_ssc,hrishietal_SEEachieva_ISIT_2024} that are useful for our purpose. We also provide some new quantities of interest in this work and present some of their properties. We state the main result of this work (Theorem \ref{thm:converse}) and present an illustrative example, comparing the converse bound and the achievable rates from \cite{hrishietal_SEEachieva_ISIT_2024}. Section \ref{sec:proofofconverse} is devoted to the proof of the converse bound. A short conclusion is provided in \ref{sec:conclusion}. The appendices contain various proofs which are not included in the main text. 

 \emph{Notation:} %In this paper, random quantities are represented as capital letters. 
 Logarithms are in base $2$. The probability of an event $E$ is denoted by $\Pr(E)$  and its complement by $\bar{E}$.  We denote strings with underlines, such as $\vx$. The notation $[m:n]$ denotes the integers $\{m,m+1,\hdots,n\}$, where $[1:n]$ is also denoted as $[n]$. The indicator variable of $E$ is denoted as $\indicatorRV_E$. 
 %Our work follows a setup similar to the \cite{ravi_coded_ssc} and inspired by the erasures caused due to quality thresholds during sequencing.

\section{Preliminaries and Main Result}
\label{section:Channel description}

\begin{figure}[tbh]
\centering
\includegraphics[width=\columnwidth]{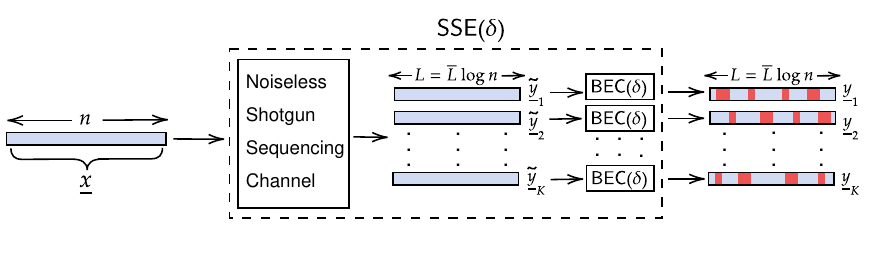}
\caption{The Shotgun Sequencing Channel with Erasures ($\ssechannel$). Red/dark colours in the output reads denote erasures.}
\label{fig:SSEchannel}
\end{figure}

%%%%%
We recall the channel model for the shotgun sequencing channel with erasures, denoted by $\ssechannel$, from \cite{hrishietal_SEEachieva_ISIT_2024}, adopting consistent notation as well. The input to the channel is an $\bl$-length string $\vx=(x_1,\hdots,x_\bl)\in \{0,1\}^\bl$, chosen from a codebook containing $2^{nR}$ codewords, where $\vx$ corresponds to the message $W \in [2^{\bl R}]$ to be transmitted. The output of the channel is given by a set $\caly$ of $\noofreads$ substrings of $\vx$ (called `reads'), where in each such substring, bits have been erased independently with probability $\erasprob$. We describe this model more precisely. The channel first samples $\noofreads$ substrings of length $L$ (in a cyclic, wrap-around fashion), denoted by  $\calynoerasures=\{\vynoerasures_1,\hdots,\vynoerasures_\noofreads\}$, where $\vynoerasures_i=(x_{\start{\vynoerasures_i}},\hdots,x_{\start{\vynoerasures_i}+\rl-1})$ is the substring starting from position $\start{\vynoerasures_i}\in[\bl]$. The collection of these starting positions are $\startpos{\cal Y}^K=\{\start{\vynoerasures_i}:i\in[\noofreads]\}$, where we assume each $\start{\vynoerasures_i}$ is chosen uniformly at random from $[\bl]$. Note that $\calynoerasures$ can be a multi-set, containing many copies of any $L$-length substring of $\vx$. Note that the collection $\calynoerasures$ can be understood as the reads obtained through a noise-free shotgun sequencing channel, following the model assumed in \cite{ravi_coded_ssc}. The channel then erases each bit in each of the noise-free reads in $\calynoerasures$ independently with probability $\delta$, to obtain the multi-set of $L$-length reads, given by $\caly=\{\vy_i\in\{0,1,\eras\}^\rl:i\in[\noofreads]\}$, where $\eras$ denotes an erasure. Note that the start positions remain unchanged, i.e., $\start{\vy_i}=\start{\vynoerasures_i}, \forall i$. Fig. \ref{fig:SSEchannel} provides an illustration.  A rate $R$ is said to be \textit{achievable} on $\ssechannel$ if the message $W$ can be reconstructed from $\caly$ using some decoding algorithm with a probability of error that is vanishing as $\bl$ grows large. The capacity of $\ssechannel$ is then defined as $C_{\ssechannel}\triangleq\sup\{R:R~\text{is achievable}\}$.

The \textit{coverage depth} of the channel model, denoted by $c$, is defined as $c\triangleq \expect(\sum_{i=1}^\noofreads\indicatorRV_{\{j\in [\start{\vy_i}:\start{\vy_i}+L-1]\}}).$  It is easy to see that $c = \frac{\noofreads\rl}{\bl},$ thus representing the expected number of times a coordinate of $\vx$ (say the $j^\tth$ coordinate) is covered by reads in $\caly$. Clearly $c>0$. As in \cite{ravi_coded_ssc, hrishietal_SEEachieva_ISIT_2024}, we operate in the regime where 
$ \rl=\normrl\log{\bl}$, 
for some positive constant $\normrl$. Thus, in our regime,
$
\noofreads=\frac{c\bl}{\normrl\log{\bl}}=\Theta\left(\frac{\bl}{\log{\bl}}\right)$. We recall a few notions, which were originally defined in \cite{ravi_coded_ssc,hrishietal_SEEachieva_ISIT_2024}.
%%%%%
\begin{definition}[\cite{ravi_coded_ssc,hrishietal_SEEachieva_ISIT_2024}] The $i^\tth$ bit of $\vx_{W}$, denoted by $x_i$, is said to be \emph{covered} by a read $\vy\in{\caly}$ if $\start{\vy} \in [i-\rl+1:i]$. Further, $x_i$ is said to be \textit{visibly covered} by $\vy$ if it is covered by $\vy$ and further unerased in $\vy$. The bit $x_i$ is said to be covered (visibly covered) by the collection of reads $\caly$ if it is covered (visibly covered, respectively) by at least one read in $\caly$.
\end{definition}
\begin{definition}
The coverage (denoted by $\Phi$) for a collection $\caly$ is the fraction of bits which are covered by reads in $\caly$, i.e, $\Phi \triangleq \frac{1}{\bl} \sum_{i=1}^{\bl} \indicatorRV_{\{ x_{i} \text{ is covered by reads in } {\caly}\}}$. The \textit{visible coverage} denoted by $\Phi_v$ of the collection $\caly$ is defined as the fraction of the bits which are visibly covered by the reads in $\caly$.  Thus, 
         $
        \Phi_v \triangleq \frac{1}{\bl} \sum_{i=1}^{\bl} \indicatorRV_{\{ x_{i} \text{ is visibly covered by reads in } {\caly}\}}.
        $
\end{definition}
%%%%%
The following are known from \cite[Lemma 2]{arxiv_Hri_SSE_24} and \cite[Lemma 1]{ravi_coded_ssc}. 
\begin{align}
\nonumber \expect[\Phi_v]&=\Pr(x_{i} \text{ is visibly covered by reads in } {\caly})\\
\label{eqn:expectedViscoverage}
&=(1-e^{-c(1-\delta)}).\\
\label{eqn:expectedCov}
\expect[\Phi]&=\Pr(x_{i} \text{ is covered by reads in } {\caly})=(1-e^{-c}).
\end{align}

We also need  the notions of merging and `true' islands, in this channel, for the purpose of proving our converse result. We recall these from \cite{hrishietal_SEEachieva_ISIT_2024,arxiv_Hri_SSE_24} briefly, for the sake of completeness. For formal definitions, we refer the reader to \cite{arxiv_Hri_SSE_24}. Two reads $\vy_i$ and $\vy_j$ of $\caly$ which have a predecessor-successor relationship in terms of their starting positions (i.e., $\vy_j$ is `nearest' to $\vy_i$ as per the starting positions) can be merged together, if they share an overlapping region (i.e., $\start{\vy_i}\leq \start{\vy_j}<\start{\vy_i}+\rl$). Note that, in the merged longer substring, we will retain any bit that is unerased in at least one of the two reads as is. Doing this successively over all reads in $\caly$ results in contiguous (erasure-containing) substrings of the input $\vx$, which are called \textit{true islands}. Here, the word `true' emphasizes the fact that the merging process is based on the knowledge of the starting positions of the reads, which is not known to the receiver in general. Let the collection of true islands be denoted by $\caly'$.  Note that, no two islands in $\caly'$ share any common read. Let $N_i:i\in[|\caly'|]$ denote the lengths of the islands in $\caly'$. Then, by the definition of coverage and islands, the following claim (originally stated in \cite[Section V, below (22)]{ravi_coded_ssc}) holds. 
\begin{align}
\label{eqn:coverageISavglengthofIslands}
    \lim_{\bl\to\infty}\frac{1}{\bl}\expect\left[\sum_{\substack{i=1}}^{|{\caly'}|}N_i\right]=\expect[\Phi]=(1-e^{-c}).
\end{align}

Note that any erasure that is found in any island in $\caly'$ is because of the fact that the bit in that position in $\vx$ was covered by the reads in $\caly$ but was erased in all the reads in $\caly$.  We have the related lemma below, the proof of which is in Appendix \ref{app:proofoflemma_islanderasures}.
%%%
\begin{lemma}
\label{lemma:islanderasureprobcalc}
Let $\delta_e$ denote the probability of a bit of $\vx$ being covered by $\caly$ but not visibly covered. Then, 
    \begin{align}
        \delta_e=\left(e^{-c(1-\erasprob)} -e^{-c} \right).
    \end{align}
\end{lemma}
%%%%
We remark that $\delta_e$ can be much smaller than $\delta$, which is intuitive, as $\delta_e$ arises out of a covered-bit not being visibly-covered in any read, unlike $\delta$ which is related to a single read. 

A similar description of merging and the formation of islands can be provided for the noise-free reads $\tilde{\caly}$. We will denote the true noise-free islands resulting from the merging of reads in $\tilde{\caly}$ as $\tilde{\caly}'$. Note that the cardinality of this collection and the lengths of each of these islands are identical to those of the collection of noise-free islands $\tilde{\caly}'$. This enables us to use some of the analysis from \cite{ravi_coded_ssc, ravi_torn_paper_lost_pieces, ravi_recovering_message_incomplete_set} as is, to aid the proofs of the results in this work. 

%%%%%
The following achievability result was shown in \cite{hrishietal_SEEachieva_ISIT_2024}.
%%%%%
\begin{theorem}
\label{thm:main}
    Let $c$ and $\normrl$ be the parameters of $\ssechannel$ such that $c>0$ and $\normrl(1-\erasprob)>1$. Let $\alpha=c/(\normrl(1-\erasprob))$. The rate $R$ is achievable on $\ssechannel$ if 
    \begin{align}
    \label{eqn:achievablerates}
     R<\left(1- e^{-c(1-\erasprob)}\right) - (1-\delta)\left(e^{-c\left(1-\frac{1}{\normrl(1-\delta)} \right)} - e^{-c}\right).
    \end{align}
\end{theorem}

In this work, we show the following converse for the capacity $C_{\ssechannel}$. 
\begin{theorem}
\label{thm:converse}
    Let $c$ and $\normrl$ be the parameters of $\ssechannel$ such that $c>0$. Let $\delta_e= \left(e^{-c(1-\erasprob)} -e^{-c} \right)$ denote the bit-erasure probability in the true islands. The capacity $C_{\ssechannel}$ of the channel $\ssechannel$ satisfies 
    \begin{align}
\label{eqn:capequal0forshortread}C_{\ssechannel}&=0,~~~\text{if}~ \normrl <\frac{c}{c-{\delta_e}},\\
    \label{eqn:capupperboundforlonger}
        C_{\ssechannel}&\leq 
        \left(1-\delta_e\right)\left(1-e^{-c}\right)-\frac{c}{\normrl}e^{-c}, ~~~\text{if}~\normrl\geq \frac{c}{c-\delta_e}. 
    \end{align}
\end{theorem}
% \begin{remark}
% Note that Theorem \ref{thm:main} implies $C_{\ssechannel}\geq \left(1- e^{-c(1-\erasprob)}\right) - \frac{ce^{-c}}{\normrl} - \upbeta(d)$, for any $d>0$. The quantity $d$ and the unwieldy expression for $\upbeta(d)$ is due to the arguments used in the proof. Simulation results show that the value of $\upbeta(d)$ decreases as $d$ reduces, and thus the expression $\upbeta(d)$ likely converges as $d\to 0$. However, we are currently unable to prove this analytically. 
% \end{remark}

% \begin{figure}[htbp]
% \centering
% \includegraphics[width=\columnwidth]{plot_legend_1.75.pdf}
% \caption{\pk{To be changed}}
% \label{fig:plot-withconverse}
% \end{figure}
% \pk{Plot and plot discussion to be changed} Fig. \ref{fig:plot} plots the upper bound for the achievable rate for $\ssechannel$ from Theorem \ref{thm:main}, for $\erasprob\in\{0,0.05,0.2,0.3\}$, against varying values for the coverage depth $c$. The parameter $\normrl$ is fixed as $1.75$ (thus satisfying the requirement $\normrl(1-\erasprob)>1$, for all chosen $\erasprob$).
%%%%
\begin{figure}[htbp]
\centering
\includegraphics[width=\columnwidth]{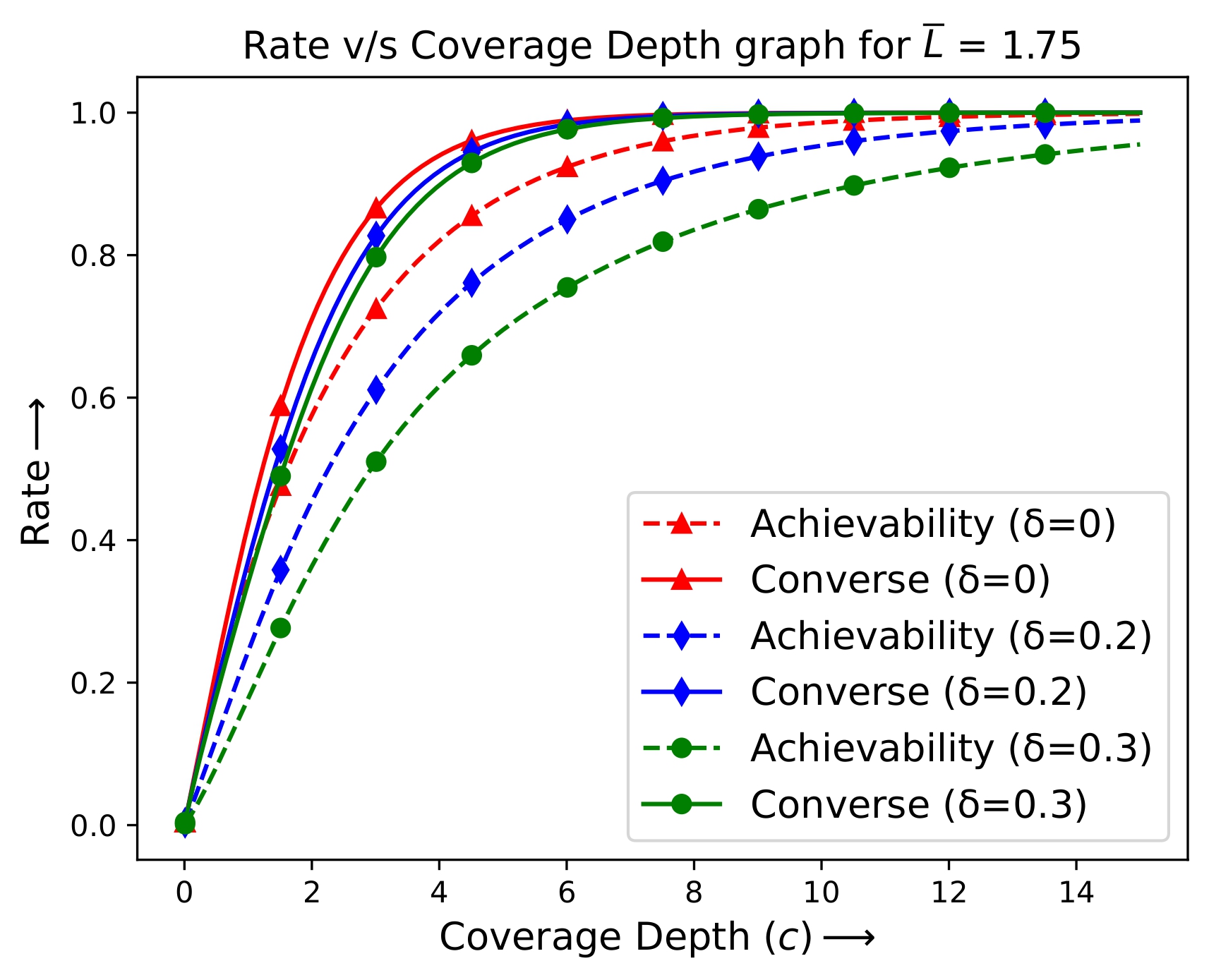}
\caption{The plot compares the achievability (Theorem \ref{thm:main}) and converse (Theorem \ref{thm:converse}) results, with $\normrl = 1.75$, for $\delta = 0, 0.2$ and $0.3$, as the coverage depth $c > 0$ varies.}
\label{fig:plot-withconverse}
\end{figure}

Fig. \ref{fig:plot-withconverse} compares the lower bound and upper bound (Theorems \ref{thm:main} and \ref{thm:converse}, respectively) on $C_{\ssechannel}$, for $\erasprob\in\{0,0.2,0.3\}$, against varying values for the coverage depth $c > 0$. The parameter $\normrl$ is fixed as $\normrl = 1.75$, to satisfy the conditions $(i)$ $\normrl \geq 1/(1-\delta)$ and $(ii)$ $\normrl \geq c/(c-\delta_e)$, for every choice of $c$ and $\delta$. The plot showcases some broad trends. Firstly, as $\delta$ increases, the gap between the achievability and converse expressions increases. Further, for extreme values of $c$, i.e., when $c$ is very low or very high, the gap reduces, eventually converging to $0$. This behaviour can also be easily observed from the expressions in Theorem \ref{thm:main} and Theorem \ref{thm:converse}. 

\section{Proof of Theorem \ref{thm:converse}}
\label{sec:proofofconverse}
In this section, we prove Theorem \ref{thm:converse}. The approach of the proof is similar to one used in \cite[Section V]{ravi_coded_ssc}. We defer the proof of \eqref{eqn:capequal0forshortread} to Appendix \ref{app:proofofshortreadcap0}, while remarking on the intuition here. Essentially, when the read lengths are small, the multi-set of reads $\caly$ have too few distinct reads, which leads to the number of distinguishable input sequences to be sub-exponential in $\bl$, thus leading to a vanishing rate. 

We now outline the arguments regarding the proof of \eqref{eqn:capupperboundforlonger}, under the assumption that $\frac{\rl}{\log\bl}=\normrl\geq \frac{c}{(c-\delta_e)},$ which we retain throughout from now in this section. The schema of the proof follows ideas from \cite{ravi_torn_paper_lost_pieces} and its expanded version \cite{ravi_recovering_message_incomplete_set}.
\begin{itemize}  
    \item Firstly, we show that capacity is essentially bounded from above by $\frac{1}{\bl}I(\vX;\caly')$, where $\vX$ is the input sequence. This corresponds to bounding the capacity via the information gathered by a `genie-aided' channel, where the output is the set of true islands ${\cal Y'}$ formed by merging the reads in ${\cal Y}$. 
    \item Via a number of careful concentration-based arguments, we show that $\frac{1}{\bl}I(\vX;\caly')$ can be upper-bounded by the difference of two terms, a `coverage' term $(1-\delta_e)\expect[\Phi]$, and a `re-ordering cost' term given by $\frac{\log\bl}{\bl}\expect[K']$, which implies a expected cost of $\log\bl$ bits of $\vX$ for each of the $K'\triangleq |\caly'|$ islands. This bound then can be proved to be equal to the bound \eqref{eqn:capupperboundforlonger} in Theorem \ref{thm:converse}.
\end{itemize}
We now present the main arguments. Due to space constraints, the proofs of many claims are provided in the appendices. 
%%%

Let $Z_{\cal Y}$ represent the collection of the erasure patterns of each of the reads in ${\cal Y}$. We now show that $H(\caly|\caly') - H(\caly|\caly', \vX)=H(\caly|Z_{\cal Y},\caly')-H(\caly|Z_{\cal Y},\caly', \vX)$. Observe that $H(Z_{\cal Y}|\caly,\caly')=0=H(Z_{\cal Y}|\caly,\caly', \vX)$, as $\caly$ determines $Z_{\cal Y}$. Also, as $\vX$ and $Z_{\cal Y}$ are independent given ${\cal Y}'$, we have that $H(Z_{\cal Y}|\caly',\vX)=H(Z_{\cal Y}|\caly')$. Thus, we have 
\begin{align}
% \label{eqn:long_read_fano_2}
    H&(\caly | \caly')= H(\caly,Z_{\cal Y}|\caly') - H(Z_{\cal Y}|\caly,\caly')\nonumber\\
    % &\stackrel{}{=} H(\caly,Z_{\cal Y}|\caly')\nonumber\\
    &=H(Z_{\cal Y}|\caly') + H(\caly|Z_{\cal Y},\caly')\nonumber\\
    &=H(Z_{\cal Y}|\caly',\vX)+ H(\caly|Z_{\cal Y},\caly')\nonumber\\
    % &=H(\caly,Z_{\cal Y}|\caly', \vX)-H(\caly|Z_{\cal Y},\caly', \vX)+ H(\caly|Z_{\cal Y},\caly')\nonumber\\
    % &=H(\caly|\caly', \vX)+H(Z_{\cal Y}|\caly, \caly', \vX)-H(\caly|Z_{\cal Y},\caly', \vX)+ H(\caly|Z_{\cal Y},\caly')\nonumber\\
    &=H(\caly|\caly', \vX)-H(\caly|Z_{\cal Y},\caly', \vX) \nonumber\\ &~~~+ H(\caly|Z_{\cal Y},\caly')\nonumber,
\end{align}
which shows that $H(\caly|\caly') - H(\caly|\caly', \vX)=H(\caly|Z_{\cal Y},\caly')-H(\caly|Z_{\cal Y},\caly', \vX)$. 
Using this, for any achievable rate $R$ on the $\ssechannel$, we get 
%%%
\begin{align}
\label{eqn:long_read_fano_1}
    &R \leq  \lim_{\bl \to \infty} \frac{I(\vX; {\cal Y})}{\bl} \leq \lim_{\bl \to \infty} \frac{I(\vX; {\cal Y},{\cal Y'})}{\bl}\nonumber\\
    &=\lim_{\bl \to \infty} \frac{I(\vX; {\cal Y'})}{\bl} + \lim_{\bl \to \infty} \frac{I(\vX; {\cal Y}|{\cal Y'})}{\bl}\nonumber\\
    &= \lim_{\bl \to \infty} \frac{I(\vX; {\cal Y'})}{\bl} + \lim_{\bl \to \infty} \frac{H(\caly|\caly') - H(\caly|\caly', \vX)}{\bl}\nonumber\\
    &\leq  \lim_{\bl \to \infty} \frac{I(\vX; {\cal Y'})}{\bl} + \lim_{\bl \to \infty} \frac{H(\caly|Z_{\cal Y},\caly')}{\bl}\nonumber\\ &~~~- \lim_{\bl \to \infty} \frac{H(\caly|Z_{\cal Y},\caly', \vX)}{\bl}\nonumber\\
    &\leq \lim_{\bl \to \infty} \frac{I(\vX; {\cal Y'})}{\bl} + \lim_{\bl \to \infty} \frac{H(\caly|Z_{\cal Y},\caly')}{\bl}.
\end{align}
%%%%%
We have the following claim, the proof of which is in Appendix \ref{app:proof_claim:hYgnZY'is0}. 
%%%%%
\begin{claim}
\label{claim:hYgnZY'is0}
    $\lim_{\bl \to \infty} \frac{H(\caly|Z_{\cal Y},\caly')}{\bl} = 0.$
\end{claim}
Using \eqref{eqn:long_read_fano_1} in conjunction with Claim \ref{claim:hYgnZY'is0}, we thus get
\begin{align}
\label{eqn:ratelessthanMIwithislands}
    R \leq \lim_{\bl \to \infty} \frac{I(\vX; {\cal Y'})}{\bl}.
\end{align}
%%%
%%%%

The rest of the proof is devoted to showing that the R.H.S. of \eqref{eqn:ratelessthanMIwithislands} can be upper-bounded by the R.H.S. of \eqref{eqn:capupperboundforlonger}. 

\subsection{A partitioning of $\caly'$ and related properties}
\label{subsec:partitionandproperties}
To do this, we partition the collection $\caly'$ into subsets and separately handle each. Towards defining this partition, we make some observations. Let $L'>0$ be some positive integer. Let $N_i$ denote the length of the $i^{\tth}$ true island in $\caly'$. Note that $N_i$ is also the length of the same island, without erasures. From \cite[Appendix G]{ravi_coded_ssc} (via an analysis applicable to islands without erasures), it follows that $\lim_{\bl\to\infty} \frac{\log\bl}{\expect[N_i]}\in(0,\infty)$, and $\lim_{\bl\to\infty} \frac{\expect[N_i^2]}{\log^2\bl}$ is finite and bounded. Further, from the assumptions regarding the channel model (specifically, the way reads are obtained), it can be seen that $\ExLenIsland\triangleq \expect[N_i]=\expect[N_1]$, as also observed from the arguments in \cite[Appendix G]{ravi_coded_ssc}. Note that $\ExLenIsland\geq L=\normrl\log\bl$, as any island must have length at least $L$. Let $F$ denote the event $\{N_i>\alpha\normrl \log^2\bl\}$, for any given $\alpha>-1/\log(1-e^{-c})$. From Lemma \ref{lemma:maxnoofreadsinislands} (see Appendix \ref{app:proof_claim:hYgnZY'is0}, originally stated in \cite[Lemma 6]{ravi_coded_ssc}), we can see that $\Pr(F)\to 0$ as $\bl\to\infty$.

For each $k$ being a non-negative integer, we define the following sets $\caly'_k$, which partition the collection of true islands $\caly'=\{\island_i:i\in[|\caly'|]\}$ into subsets containing islands of approximately equal lengths. 
%%%%
\begin{align}
    \label{eqn:defnofyk'}
    \caly'_k \triangleq \left\{\island_i \in \caly' \,:\, \left(\frac{k-1}{L'}\right)\log \bl \leq N_i < \frac{k}{L'}\log \bl\right\}
\end{align}
We further denote $\caly'_{\geq J} \triangleq \bigcup_{k\geq J}\caly'_k $.

 We recall arguments from  \cite[Lemma 3]{ravi_recovering_message_incomplete_set} which show that the cardinality $|\caly'_k|$ concentrates around its mean.  Towards that end, we define the term $q_{k,\bl}$, following notations in \cite{ravi_recovering_message_incomplete_set}, as $
 % \begin{align*}
    q_{k,\bl} \triangleq \Pr\left(\left(\frac{k-1}{L'}\right)\log{\bl} \leq N_i < \frac{k}{L'}\log{\bl}\right).
    % \ekn &= \Pr\left(\island_i \in \caly' \mid \left(\frac{k}{L'}-\frac{1}{1-\delta_e}\right)\log{\bl} \leq N_i <\frac{k}{L'}\log{\bl}\right)
% \end{align*}
%%%
$
Observe that the above probability is unchanged with the index $i$, as the $N_i$ is identically distributed for all $i\in[|\caly'|]$. By definition, we see that, by the fact that $N_i\geq L=\normrl\log\bl$, we have $q_{k,\bl}\ekn=0,$ for $k<\frac{L'+1}{1-\delta_e}$. Thus, we see that, for any $i\in[|\caly'|]$, for $J\geq 2$ being a positive integer,
%%%%
\begin{align}
\label{eq:sumofprobabilitiesofIslands}\lim_{J\to\infty}\sum_{k=1}^{J-1}q_{k,n}\ekn=\sum_{k=\frac{L'+1}{1-\delta_e}}^{J-1}q_{k,\bl}\ekn= \Pr(\island_i\in \caly')=1.
\end{align}    

From \cite[Proof of Lemma 3]{ravi_recovering_message_incomplete_set}, it is seen that $\expect[|\caly'_k|]=\frac{\bl q_{k,\bl}\ekn}{\ExLenIsland}$. For any constant $\epsilon>0$, let the event ${\cal E}_{k,\bl}$ be defined as 
\begin{align}
    {\cal E}_{k,\bl} = \left\{\left||\caly'_k|-\frac{\bl q_{k,\bl}\ekn}{\ExLenIsland}\right| > \epsilon \frac{\bl}{\ExLenIsland}\right\}.
    \label{eq:defnofCalEkn}
\end{align}
%%%%
Further, we have the concentration  \cite[Proof of Lemma 3]{ravi_recovering_message_incomplete_set}.
%%%
\begin{align}
\label{eqn:probofcalEgoesto0}
\Pr({\cal E}_{k,\bl}) = O\left(e^{-\frac{\bl}{2\ExLenIsland}\epsilon^2}\right). 
\end{align}
%%%%%
Observe that $\Pr({\cal E}_{k,\bl})\to 0$ as $n\to \infty$, for any 
% \begin{align}
% \label{eqn:minepsiloncondition}
$\epsilon\geq \log\bl\left(\sqrt{\frac{\bl}{2\ExLenIsland}}\right)^{-1}$, which we will assume to be true throughout this work.
%$\epsilon$ satisfies the bound \eqref{eqn:minepsiloncondition} throughout.
We note that, under the event $\bar{\cal E}_{k,\bl}$, using \eqref{eq:defnofCalEkn}, we have
%%%%
\begin{align}
    \frac{\bl}{\ExLenIsland}( q_{k,\bl}\ekn - \epsilon) \leq |\caly'_k| \leq \frac{\bl}{\ExLenIsland}( q_{k,\bl}\ekn + \epsilon). 
\label{eq:boundonsizeofyk'} 
\end{align}
%%%%
\subsection{Proving \eqref{eqn:capupperboundforlonger}}
We now establish \eqref{eqn:capupperboundforlonger} using \eqref{eqn:ratelessthanMIwithislands} and the properties established in Subsection \ref{subsec:partitionandproperties}, via a series of claims. 

Let $\Gamma^\bl\subseteq \{0,1,2\}^\bl$ be a $n$-tuple defined as follows: the $i^{\tth}$ position in the tuple has (a) a $0$, if it is neither a starting nor an ending location of an island in $\caly'$, (b) a $1$, if an island in ${\cal Y}'$ starts at that location (reading from left to right), and (c) a $2$, if an island ends at that location.  We then have the following straightforward inequalities, for any positive integer $J\geq 2$
%%%
\begin{align}
I(\vX;\caly')&= H(\caly') - H(\caly'|\vX) \nonumber\\
\label{eqn:initialineqMI} &\overset{}{\leq} \Sigma_{k=1}^{J-1}H(\caly'_{k}) + H(\caly'_{\geq J}) - H(\caly'|\vX,\Gamma^\bl).
\end{align}
Let $Z_{\caly'}$ be the collection of erasure patterns of the set of islands $\caly'$. That is, $Z_{\caly'}$ contains precisely $|\caly'|$ binary strings, each of length corresponding to its respective island in $\caly'$, where in any such string a $0$ at a position indicates a non-erasure at that position, whereas a $1$ indicates an erasure at the same position. Similarly, let $Z_{\caly'_k}$ denote the set of error patterns corresponding to islands in $\caly'_k$. Observe that $H(Z_{\caly'}|\vX,\Gamma^\bl,\caly') = 0$, as we have the set of islands $\caly'$ from which the corresponding erasure patterns can be obtained. Further, $H(\caly'|\vX,\Gamma^\bl,Z_{\caly'})=0$ as well, since using $\vX,\Gamma^\bl,Z_{\caly'}$ we can reconstruct all islands in $\caly'$. Using these observations, we obtain the following.
\begin{align}
H&(\caly'|\vX,\Gamma^\bl) = H(\caly',Z_{\caly'}|\vX,\Gamma^\bl) - H(Z_{\caly'}|\vX,\Gamma^\bl,\caly') \nonumber\\
    &= H(Z_{\caly'}|\vX,\Gamma^\bl) + H(\caly'|\vX,\Gamma^\bl,Z_{\caly'}){=} H(Z_{\caly'}|\vX,\Gamma^\bl)\nonumber\\
    &\stackrel{(a)}{\geq} \sum_{k=1}^{J-1}H(Z_{\caly'_k}|\vX,\Gamma^\bl) \geq \sum_{k=1}^{J-1}H(Z_{\caly'_k}|\vX,\Gamma^\bl,\bar{\cal E}_{k,\bl})
    \label{eqref:16},
\end{align}
where $(a)$ holds as the erasures in different islands are independent. 
% We thus obtain
% \begin{align}
%     I(\vX;\caly')&\leq \sum_{k=1}^{J-1}H(\caly'_k) + H(\caly'_{\geq J}) - H(Z_{\caly'}|\vX,\Gamma^\bl).
%     \end{align}

% \pk{The upper limit for $k$ is $nL/\log \bl$} 
% \pk{$\Tilde{}$ notation for complement or something else? This comes in $\bar{\cal E}_{k,\bl}$ as well as other events, so please check. }

Thus, using \eqref{eqref:16} in \eqref{eqn:initialineqMI}
\begin{align}\label{eqn:reducedmutualinfoislandsandinput}
    &\frac{1}{\bl}I(\vX;\caly')\nonumber \\
    &\leq \frac{1}{\bl}\sum_{k=1}^{J-1}\left(H(\caly'_k)- H(Z_{\caly'_k}|\vX,\Gamma^\bl,\bar{\cal E}_{k,\bl})\right) + \frac{1}{\bl}H(\caly'_{\geq J}). 
    \end{align}
In the following two claims, we bound the first term in the R.H.S of \eqref{eqn:reducedmutualinfoislandsandinput}. The proofs are in Appendix \ref{proof:claim:bounonHcalyk'byn} and Appendix \ref{app:proofoffinalclaim} respectively. 
%%%%%%
\begin{claim}
\label{claim:bounonHcalyk'byn} For $J\in[2:\beta\log\bl]$ (for any $\beta>0$ being a constant), we have
     \begin{align*}
        \frac{1}{\bl}&\sum_{k=1}^{J-1}H(\caly'_k)
        \leq \frac{1}{\bl}(1-\delta_e)\sum_{k=1}^{J-1}\expect\left[\sum_{\substack{i=1}}^{|\caly'|}N_i~|~~\bar{\cal E}_{k,\bl}\right]\nonumber\\
  &+\sum_{k=1}^{J-1}\frac{q_{k,\bl}\ekn}{\ExLenIsland}\left(\log e -\log\bl- \log\left(\frac{q_{k,n}\ekn -\epsilon}{\ExLenIsland}\right)\right)\nonumber\\
  &~~~+ \frac{1}{\bl}\sum_{k=1}^{J-1}H(Z_{\caly'_k}|\vX,\Gamma^\bl,\bar{\cal E}_{k,\bl})+o(1).
\end{align*}
\end{claim}
%%%%%
Using Claim \ref{claim:bounonHcalyk'byn}, we can prove the following. 
\begin{claim}
    \label{claim:final}
   For $J \in [2:\beta\log\bl]$ (for any $\beta>0$ being a constant), we have
    \begin{align*}
    &\lim_{\bl\to \infty}\left(\sum_{k=1}^{J-1}\frac{H(\caly'_k)}{\bl}-\sum_{k=1}^{J-1} \frac{H(Z_{\caly'_k}|\vX,\Gamma^\bl,\bar{\cal E}_{k,\bl})}{\bl} \right)\nonumber\\
    &\stackrel{}{\leq} \left(1-\delta_e\right)\lim_{\bl\to\infty}\frac{1}{\bl}\expect\left[\sum_{\substack{i=1}}^{|{\caly'}|}N_i\right] - \lim_{\bl\to \infty}\frac{\log\bl }{\bl}\expect[\noofreads']\sum_{k=1}^{J-1}q_{k,\bl}\ekn
    \end{align*}
\end{claim}
%%%

We bound the second term in the R.H.S. of \eqref{eqn:reducedmutualinfoislandsandinput} via the following claim, which is almost identical to \cite[Lemma 4]{ravi_recovering_message_incomplete_set}. The proof of the same is in Appendix \ref{app:proof:claim:entropyofy'GreaterthanJgoesto0}.
%%%%
\begin{claim}
%%%%
\label{claim:entropyofy'GreaterthanJgoesto0}
    For every $J,L'$ and $\zeta>0$, the entropy of  the set $\caly'_{\geq J}$ is bounded from above as
    %%%
    \begin{align}
    \frac{1}{\bl}H(\caly'_{\geq J})\leq 3(S\sqrt{L'/(J-1)} + \zeta) + o(1),
    \end{align}
    for some finite $S$.
\end{claim} 

Finally, we can get our converse in Theorem \ref{thm:converse} by putting together \eqref{eqn:reducedmutualinfoislandsandinput} with Claims \ref{claim:final} and \ref{claim:entropyofy'GreaterthanJgoesto0}. To do this, we let $J$ grow large (as fast as $\beta\log\bl$, for suitable constant $\beta$). This ensures $\frac{1}{\bl}H(\caly'_{\geq J})\to 0$, by Claim \ref{claim:entropyofy'GreaterthanJgoesto0}. At the same time, we have
\begin{align*}
\lim_{\bl\to \infty}\frac{\log\bl }{\bl}\expect[\noofreads']\sum_{k=1}^{J-1}q_{k,\bl}\ekn&\stackrel{(a)}{=} \lim_{\bl\to \infty}\frac{\log\bl }{\bl}\expect[\noofreads']\\
&\stackrel{(b)}{=}\lim_{\bl\to \infty}\frac{\log\bl }{\bl}\noofreads e^{-c}\stackrel{}{=}\frac{c}{\normrl}e^{-c},
\end{align*}
where $(a)$ holds as all islands have lengths $\Theta(\log^2\bl)$, by observations in Sub-section \ref{subsec:partitionandproperties}, and $(b)$ holds by \cite[Lemma 2]{ravi_coded_ssc}. Finally, observe that $\lim_{\bl\to\infty}\frac{1}{\bl}\expect\left[\sum_{\substack{i=1}}^{|{\caly'}|}N_i\right]=(1-e^{-c})$, by \eqref{eqn:coverageISavglengthofIslands}. 
\hspace{-0.2cm}
\section{Conclusion}
\label{sec:conclusion}
We have presented a converse for the shotgun sequencing channel with erasures. While this is not tight, we have attempted to provide a self-contained analysis that can possibly be sharpened to obtain a tight converse, similar to that in \cite{ravi_coded_ssc}. Finding the capacity of shotgun sequencing channel under other noise models is also an important line of work to be addressed. 
\clearpage
\IEEEtriggeratref{10}
\bibliographystyle{IEEEtran}
\bibliography{ISIT24_ShotgunErasures_ArXiv.bib}
%%

%%%%%
\clearpage
\appendices
\section{Proof of Lemma \ref{lemma:islanderasureprobcalc}}
\label{app:proofoflemma_islanderasures}
Let $E_c$ and $E_v$ be the events that bit $x_1$ in the input sequence $\vx$ is covered and visibly covered respectively. Hence the required probability is as follows.
\begin{align}
\label{eqn:probabilityofnotvisiblycoveredANDcovered}
\nonumber
     \Pr(x_1 &\text{ is covered but not visibly covered by reads in } {\cal Y})\\
     \nonumber
     &=\Pr(E_c,\bar{E_v})\\
     &= \Pr(\bar{E_v}\vert E_c)\Pr(E_c) = (1- \Pr(E_v\vert E_c)) \Pr(E_c)
    \nonumber\\
    \nonumber
    &\stackrel{(a)}{=}\left(1- \frac{\Pr(E_v)}{\Pr(E_c)}\right) \Pr(E_c) = \Pr(E_c) - \Pr(E_v)\\
    &\stackrel{(b)}{=} (1-e^{-c}) - (1-e^{-c(1-\erasprob)}) = \left(e^{-c(1-\erasprob)} -e^{-c} \right),
\end{align}
where $(a)$ follows by the Bayes' rule and since $\Pr(E_c, E_v)=\Pr(E_v)$, and $(b)$ follows from \eqref{eqn:expectedViscoverage} and \eqref{eqn:expectedCov}.
%%%
\section{{Proof of \eqref{eqn:capequal0forshortread}}}
\label{app:proofofshortreadcap0}

We follow arguments along the lines of those in \cite[Section V]{ravi_coded_ssc}. Let $\vX$ be the random variable representing the $n$-length input to the channel. Recall that ${\cal{Y}}$ and ${\cal \tilde{Y}}$ are the sets of reads with and without erasures respectively. Any achievable rate $R$ on $\ssechannel$ satisfies the following.

\begin{align}
\label{eqn:converse_fano}
    R &\leq \lim_{\bl \to \infty} \max_{P_{{\vX}}} \frac{I({\vX}; {\cal{Y}})}{\bl}\nonumber\\
        &\stackrel{(a)}{\leq} \lim_{\bl \to \infty} \max_{P_{{\vX}}} \frac{I({\cal \tilde{Y}}; {\cal{Y}})}{\bl}\nonumber\\
        &= \lim_{\bl \to \infty} \max_{P_{{\vX}}} \frac{H({\cal \tilde{Y}}) - H({\cal \tilde{Y}}|{\cal{Y}})}{n}\nonumber\\
        &\stackrel{(b)}{\leq} \lim_{\bl \to \infty} \left(\frac{\noofreads \log{(1 + \frac{2^\rl - 1}{\noofreads})}}{\bl} - \max_{P_{{\vX}}}\frac{ H({\cal \tilde{Y}}|{\cal{Y}})}{\bl}\right),
\end{align}

where (a) is due to the data processing inequality and (b) follows from the arguments in Section V in \cite[Section V,~`Short reads']{ravi_coded_ssc}. 
% \pk{I had some other argument for (b) in my notes, to avoid using \cite{ravi_coded_ssc}. Do check} 

Recall that ${\cal Y}'$ and ${\cal \tilde{Y}}'$ denote the sets of true islands formed from ${\cal Y}$ and ${\cal \tilde{Y}}$, respectively.
Let $N_e({\cal Y}')$ be the number of bits in ${\cal Y}'$ that remain erased after merging $\caly$ to form $\caly'$. Now, we have
%%%%
\begin{align}
\label{eqn:entropy_of_tilda_Y_given_Y}
    H({\cal \tilde{Y}}|{\cal Y})&\stackrel{(a)}{\geq}H\left({\cal \tilde{Y}}|{\cal Y},{\cal Y}',\startpos{\cal Y}^K\right)\nonumber\\
    &\stackrel{}{=}H\left({\cal \tilde{Y}}|{\cal Y}',\startpos{\cal Y}^K\right)-I\left({\cal \tilde{Y}};{\cal Y}|{\cal Y}',\startpos{\cal Y}^K\right)\nonumber\\
    &\stackrel{(b)}{=}H\left({\cal \tilde{Y}}|{\cal Y}',\startpos{\cal Y}^K\right)\nonumber\\
    &\stackrel{}{=}H\left({\cal \tilde{Y}},{\cal \tilde{Y}}'|{\cal Y}',\startpos{\cal Y}^K\right)-H\left({\cal \tilde{Y}}'|{\cal \tilde{Y}},{\cal Y}',\startpos{\cal Y}^K\right)\nonumber\\
    &\stackrel{(c)}{=}H\left({\cal \tilde{Y}}'|{\cal Y}',\startpos{\cal Y}^K\right)+H\left({\cal \tilde{Y}}|{\cal \tilde{Y}}',{\cal Y}',\startpos{\cal Y}^K\right)\nonumber\\
    &\stackrel{(d)}{=}H\left({\cal \tilde{Y}}'|{\cal Y}',\startpos{\cal Y}^K\right)\nonumber\\
    &\stackrel{}{=}H\left({\cal \tilde{Y}}'|{\cal Y}'\right)-I\left({\cal \tilde{Y}}';\startpos{\cal Y}^K|{\cal Y}'\right)\nonumber\\
    &\stackrel{(e)}{=}H\left({\cal \tilde{Y}}'|{\cal Y}'\right)\nonumber\\
    &\stackrel{}{=}\sum_{{\B y}'}P_{{\cal Y}'}\left({\B y}'\right)H\left({\cal \tilde{Y}}'|{\cal Y}'={\B y}'\right)\nonumber\\
    &\stackrel{(f)}{=}\expect\left[N_e\left({\cal Y}'\right)\right]\nonumber\\
    &\stackrel{(g)}{=}\expect \left[ \sum_{i=1}^{\bl} \indicatorRV_{\{ x_{i} \text{ is covered but not visibly covered by in reads in } {\caly}\}} \right]\nonumber\\
    &\stackrel{}{=}\bl\delta_e,
    % &\stackrel{}{=}\text{To-be-shown to be equal to }\frac{KL}{n}(e^{-c(1-\delta)}-e^{-c})
\end{align}
 % where $N_e({\cal Y}')$ is the number of bits in ${\cal Y}'$ that remain erased after the merging.
 Here, $(a)$ holds as conditioning reduces entropy, $(b)$ holds as no further information about ${\cal \tilde{Y}}$ is obtained from ${\cal Y}$ once ${\cal Y}'$ and $\startpos{\cal Y}^K$ together are known, $(c)$ holds since ${\cal \tilde{Y}}'$ is obtained deterministically from ${\cal \tilde{Y}}$ and $\startpos{\cal Y}^K$, $(d)$ holds since ${\cal \tilde{Y}}$ is known given ${\cal \tilde{Y}}'$ and $\startpos{\cal Y}^K$, $(e)$ is true since $I({\cal \tilde{Y}}';\startpos{\cal Y}^K|{\cal Y}')=0$ as no further information about $\startpos{\cal Y}^K$ is garnered from ${\cal \tilde{Y}}'$ after knowing ${\cal Y}'$, 
 % and finally 
 $(f)$ holds since $H({\cal \tilde{Y}}'|{\cal Y}'={\B y}')$ is precisely the number of erased bits in ${\cal \tilde{Y}}'$, since each such bit is generated as per the uniform distribution, and finally $(g)$ holds as the number of erased bits in ${\cal \tilde{Y}}'$ are equal to the number of bits in $\vX$ that are covered but not visibly covered by reads in ${\cal Y}$.

Using Lemma \ref{lemma:islanderasureprobcalc} in \eqref{eqn:entropy_of_tilda_Y_given_Y}, and this result subsequently in \eqref{eqn:converse_fano}, 
% and (\ref{eqn:expectation_of_Ne_calculation}), 
we have,
\begin{align*}
    R &\leq \lim_{\bl \to \infty} \left(\frac{\noofreads \log{(1 + \frac{2^\rl}{\noofreads})}}{\bl} - \frac{\bl(e^{-c(1-\delta)}- e^{-c})}{\bl}\right)\\
        &= \lim_{\bl \to \infty} \frac{\noofreads}{\bl}\left(\log{\left(1 + \frac{2^\rl}{\noofreads}\right)} - \frac{\rl}{\cover}(e^{-c(1-\delta)}- e^{-c})\right)\\
         &\leq \lim_{\bl \to \infty} \frac{\noofreads}{\bl}\log{\left(1 + \frac{2^\rl}{\noofreads\cdot2^{ \frac{\rl}{\cover}(e^{-c(1-\delta)}- e^{-c})}}\right)}\\
         &\stackrel{(a)}{\leq}\lim_{\bl \to \infty} \frac{2^{ \rl\left(1- \left( (e^{-c(1-\delta)}- e^{-c})/c\right)\right)}}{\bl} + {\bigo}(\frac{1}{\bl})\\ 
        &= \lim_{\bl \to \infty} \bl^{\normrl \left(1- \left((e^{-c(1-\delta)}- e^{-c})/c\right)\right)-1},
\end{align*}
where $(a)$ holds as $\log(x+1) \leq x, \forall x > -1$. Thus, if $\normrl\left(1- \left((e^{-c(1-\delta)}- e^{-c})/c\right)\right) < 1$, then the capacity decays to $0$.
%%%%%%
\section{Proof of Claim \ref{claim:hYgnZY'is0}}
\label{app:proof_claim:hYgnZY'is0}
Note that the quantity $H\left({\cal Y}| Z_{\caly},{\cal Y'}\right)$, corresponds to the randomness in the possible ways to locate the $K$ reads in ${\cal Y}$ along the islands in ${\cal Y}'$, in such a way that the reads (after incorporating erasures as per the patterns in $Z_{\cal Y}$), align to give ${\cal Y}'$. We may thus bound the quantity $H(\caly|Z_{\cal Y},\caly')$ by the number of ways to choose $K$ $L$-length substrings from the islands $\caly'$. To do this, we first bound the total number of $L$-length substrings that can be chosen from $\caly'$, and then bound the number of ways to choose $K$ substrings from this collection. This technique is used in \cite[Section V]{ravi_coded_ssc}, which we adopt here with necessary changes. 

%To obtain such a bound, we first calculate a bound on the length of each true island, for which we need the maximum number of reads that merge together to form any true island. To analyse this value, we first upper-bound on the number of reads in true islands formed by the genie. 

Let the  maximum number of reads involved in any true island in $\caly'$ be $D$. Since we are concerned with true islands here, which are unaffected by erasures, the following lemma  holds directly from \cite[Lemma 6]{ravi_coded_ssc}.
\vspace*{1em}

%%%
\begin{lemma} \cite[Lemma 6]{ravi_coded_ssc}
\label{lemma:maxnoofreadsinislands}
    For any $\alpha>-1/\log{\left(1-e^{-c}\right)}$, we have that $\Pr(D>\alpha \log{\bl}) \to 0$ as $n\to \infty.$
\end{lemma}
% \pk{$B$ seems to be unnecessary} Recall the event $B$ as defined in \eqref{eqn:defnB} in Sub-section \ref{subsec:preiciseanalysisofdecoding}. 
Thus, for a fixed $\alpha>-1/\log{\left(1-e^{-c}\right)}$, we have,
\begin{align}
\label{eqn:read_entropy_given_era_pat_and_islands}
\nonumber
     &\lim_{\bl \to \infty} \frac{H({\cal Y}|Z_{\caly},{\cal Y'})}{\bl} \leq \lim_{\bl \to \infty} \frac{1}{\bl}H\left({\cal Y}, \indicatorRV_{\{D\leq \alpha \log{\bl}\}}| Z_{\caly},{\cal Y'}\right)\\
     &\leq \lim_{\bl \to \infty}\frac{1}{\bl}\left(1 +  H\left({\cal Y}| Z_{\caly},{\cal Y'}, \indicatorRV_{\{D\leq \alpha \log{\bl}\}}\right)\right)\nonumber\\
      &\leq \lim_{\bl \to \infty} \frac{1}{\bl} H\left({\cal Y}| Z_{\caly},{\cal Y'},\{D\leq \alpha \log{\bl}\}\right)\nonumber\\\nonumber
      &~~~~+\big(\lim_{\bl \to \infty} \frac{1}{\bl} H\left({\cal Y}| Z_{\caly},{\cal Y'},\{D > \alpha \log{\bl}\}\right)\\\nonumber
      &~~~~~~~~~~~~~~~~~~~~~~~~\cdot\left(\Pr{\left( D > \alpha \log{\bl}\right)}\right)\nonumber\big)\\
      &\nonumber \stackrel{(a)}{\leq} \lim_{\bl \to \infty} \frac{1}{\bl} H\left({\cal Y}| Z_{\caly},{\cal Y'},\{D\leq \alpha \log{\bl}\}\right) \\\nonumber &~~~~~+ \lim_{\bl \to \infty} \frac{1}{\bl} \left(2\bl\right)\left(\Pr{\left( D > \alpha \log{\bl}\right)}\right)\nonumber\\
      &=\lim_{\bl \to \infty} \frac{1}{\bl} H\left({\cal Y}| Z_{\caly},{\cal Y'},\{D\leq \alpha \log{\bl}\}\right),
\end{align}
% THIS BELOW VERSION HAS event B, which I think is not necessary. \begin{align}
% \label{eqn:read_entropy_given_era_pat_and_islands}
%      \lim_{\bl \to \infty}& \frac{H({\cal Y}|Z_{\caly},{\cal Y'})}{\bl} \leq \lim_{\bl \to \infty} \frac{1}{\bl}H\left({\cal Y}, \indicatorRV_{\bar{B},\{D\leq \alpha \log{\bl}\}}| Z_{\caly},{\cal Y'}\right)\\
%      &\leq \lim_{\bl \to \infty}\frac{1}{\bl}\left(1 +  H\left({\cal Y}| Z_{\caly},{\cal Y'}, \indicatorRV_{\bar{B},\{D\leq \alpha \log{\bl}\}}\right)\right)\nonumber\\
%       &\leq \lim_{\bl \to \infty} \frac{1}{\bl} H\left({\cal Y}| Z_{\caly},{\cal Y'},\bar{B},\{D\leq \alpha \log{\bl}\}\right)\nonumber\\\nonumber
%       &~~~~+\big(\lim_{\bl \to \infty} \frac{1}{\bl} H\left({\cal Y}| Z_{\caly},{\cal Y'},B \cup \{D > \alpha \log{\bl}\}\right)\\\nonumber
%       &~~~~~~~~~~~~~~~~~\cdot\left(\Pr{\left(B\right)}+\Pr{\left( D > \alpha \log{\bl}\right)}\right)\nonumber\big)\\
%       &\nonumber \stackrel{(a)}{\leq} \lim_{\bl \to \infty} \frac{1}{\bl} H\left({\cal Y}| Z_{\caly},{\cal Y'},\bar{B},\{D\leq \alpha \log{\bl}\}\right) \\\nonumber &~~~~~+ \lim_{\bl \to \infty} \frac{1}{\bl} \left(2\bl\right)\left(\Pr{\left(B\right)}+\Pr{\left( D > \alpha \log{\bl}\right)}\right)\nonumber\\
%       &=\lim_{\bl \to \infty} \frac{1}{\bl} H\left({\cal Y}| Z_{\caly},{\cal Y'},\bar{B},\{D\leq \alpha \log{\bl}\}\right),
% \end{align}
where (a) holds because of the following reasons: (i) ${\cal Y}$, given $Z_{\cal Y}$, is completely determined by $\vX$ and $\startpos{\cal Y}^K$ (which totally have entropy at most $2n$), and (ii) since, from Lemma \ref{lemma:maxnoofreadsinislands}, we have $\lim_{\bl\to\infty}\Pr(D>\alpha\log\bl)=0$.

To bound 
%$H\left({\cal Y}| Z_{\caly},{\cal Y'},\bar{B},\{D\leq \alpha \log{\bl}\}\right)$
$H\left({\cal Y}| Z_{\caly},{\cal Y'},\{D\leq \alpha \log{\bl}\}\right)$, we now bound the number of ways to locate the reads in ${\cal Y}$ along the islands in ${\cal Y}'$ as follows, given that the event
%$\bar{B},
$\{D\leq \alpha \log{\bl}\}$ holds.  Given that the maximum number of reads per island $D\leq \alpha \log{\bl}$, total length of an island is at most $\rl\cdot \alpha \log{\bl}$, which is also an upper-bound for the total number of substrings (of any length) per island. Thus, as the total number of islands $\noofreads' < \noofreads$, the total number $\rl$ substrings in the islands of $\cal Y'$ is at most $\noofreads\rl\cdot \alpha \log{\bl} = \cover\bl\alpha\log{\bl}$. 

%\hn{[25.03.2025] As all selections of $\caly$ are not allowed, the current calculation is an upperbound. Some explaination might be needed for this.}

The term 
%$H\left({\cal Y}| Z_{\caly},{\cal Y'},\bar{B},\{D\leq \alpha \log{\bl}\}\right)$
$H\left({\cal Y}| Z_{\caly},{\cal Y'},\{D\leq \alpha \log{\bl}\}\right)$, thus, can be thought of as choosing $K$ substrings over all the possible $\rl$-length substrings of the set of islands $\caly '$. From (\ref{eqn:read_entropy_given_era_pat_and_islands}), we thus 
\begin{align}
\label{eqn:long_read_H(Y|Y',Zy,B,D)->0}
    \lim_{\bl \to \infty} \frac{H({\cal Y}|Z_{\caly},{\cal Y'})}{\bl} &\leq\lim_{\bl \to \infty} \frac{1}{\bl} H\left({\cal Y}| Z_{\caly},{\cal Y'},\{D\leq \alpha \log{\bl}\}\right)\nonumber\\
    &\leq \lim_{\bl \to \infty} \frac{1}{\bl}\log\left(\frac{(c\bl\alpha\log\bl)^\noofreads}{\noofreads !}\right)\nonumber\\
    &\leq \lim_{\bl \to \infty} \frac{1}{\bl} K\log\left(\frac{ec\bl\alpha\log\bl}{\noofreads}\right)\nonumber\\
    &\leq \lim_{\bl \to \infty} \frac{1}{\bl}\cdot \left(\frac{\cover\bl}{\log{\bl}}\log\left(e\normrl\alpha\log^2{\bl}\right)\right)\nonumber\\
    &= 0. 
\end{align}
This completes the proof of the claim.
%%%%
\section{Proof of Claim \ref{claim:bounonHcalyk'byn}}
\label{proof:claim:bounonHcalyk'byn}
Recalling the definition of the event ${\cal E}_{k,\bl}$ in \eqref{eq:defnofCalEkn} and by \eqref{eqn:probofcalEgoesto0}, we bound $ H(\caly'_k)$ as follows.

\begin{align}
    \nonumber H(\caly'_k) &\leq H(\caly'_k , \indicatorRV_{{\cal E}_{k,\bl}}) = H(\indicatorRV_{{\cal E}_{k,\bl}}) + H(\caly'_k |\indicatorRV_{{\cal E}_{k,\bl}}) \\
      \nonumber &\leq 1 + H(\caly'_k |\indicatorRV_{{\cal E}_{k,\bl}}) \leq 1 + 3n \Pr({\cal E}_{k,\bl}) + H(\caly'_k |\bar{\cal E}_{k,\bl})\\
      &\leq H(\caly'_k |\bar{\cal E}_{k,\bl})+o(1)
      \label{eqn:initialboundonEntropyofcalyk'}
\end{align}
% \pk{We have to recall that $Pr({\cal E}_{k,\bl})$ decays exponentially with $\bl$}
% \pk{March 3: Here too, $H(\caly'_k |{\cal E}_{k,\bl})\leq (2n + H_2(\delta)\bl)$ can be replaced with $H(\caly'_k |{\cal E}_{k,\bl})\leq 3n$, which is a looser but nothing-to-lose bound, in the case of SSE?}

%  Similar to \eqref{eq:11}, we can prove that the second sum becomes small as the value of J gets high.

% \begin{align}
% H(z_{\caly'_k}|X^\bl,T^\bl) \geq H(z_{\caly'_k}|X^\bl,T^\bl,\bar{\cal E}_{k,\bl})) 
% \end{align}

%  Using \eqref{eqref:16} in 

We now bound $H(\caly'_k |\bar{\cal E}_{k,\bl})$. Recall that $Z_{\caly'_k}$ denote the erasure patterns for all islands in $\caly'_k$. Then, 
\begin{align}
\nonumber
    H(\caly'_k |\bar{\cal E}_{k,\bl})   &\leq H(\caly'_k,Z_{\caly'_k}|\bar{\cal E}_{k,\bl})\\
&=H(\caly'_k|Z_{\caly'_k},\bar{\cal E}_{k,\bl}) + H(Z_{\caly'_k}|\bar{\cal E}_{k,\bl}),
    \label{eq:11}
    \end{align}
where
\begin{align}
\label{eqn:expansionofyk'givenZyk'andScriptEtilde}
    \nonumber H&(\caly'_k|Z_{\caly'_k},\bar{\cal E}_{k,\bl}) \\
    &= \sum_{z_{\caly'_k}}P(Z_{\caly'_k} = z_{\caly'_k} | \bar{\cal E}_{k,\bl})H(\caly'_k|Z_{\caly'_k} = z_{\caly'_k},\bar{\cal E}_{k,\bl}).
\end{align}
Now, $z_{\caly'_k}$ represents a candidate set of erasure patterns for $\caly'_k$. Under the condition that $Z_{\caly'_k} = z_{\caly'_k}$, let $l_{k,1},l_{k,2},....,l_{k,|\caly'_k|}$ be the length of each read in $\caly'_k$. Similarly, let $\rho_{k,1},\rho_{k,2},....,\rho_{k,|\caly'_k|}$ be the number of erasures in each read in $\caly'_k$ when $Z_{\caly'_k} = z_{\caly'_k}$. Note that both these collections (of lengths of the islands in $\caly'_k$, and of the number of erasures in islands in $\caly'_k$) are known, given the collection of erasure patterns $Z_{\caly'_k} = z_{\caly'_k}$.   Using these observations, we have, by simple counting arguments,
\begin{align}
    &2^{H(\caly'_k|Z_{\caly'_k} = z_{\caly'_k},\bar{\cal E}_{k,\bl})} \nonumber \\
    &\leq \frac{\displaystyle \prod_{j=1}^{|z_{\caly'_k}|} 2^{l_{k,j} - \rho_{k,j}}}{|z_{\caly'_k}|!} = \frac{\displaystyle \prod_{j=1}^{|z_{\caly'_k}|} 2^{l_{k,j}} \displaystyle \prod_{j=1}^{|z_{\caly'_k}|} 2^{- \rho_{k,j}}}{|z_{\caly'_k}|!},
    \label{eq:17}
\end{align}
%%%%
where, the denominator-term $|z_{\caly'_k}|!$ occurs because $\caly'_k$ is a unordered collection of $|z_{\caly'_k}|!$ islands, of respective lengths $l_{k,j}:\forall j$. Now the term $\frac{\displaystyle \prod_{j=1}^{|z_{\caly'_k}|} 2^{l_{k,j}}}{|z_{\caly'_k}|!}$ can be bounded as follows. 
% \begin{align*}
% \frac{\displaystyle \prod_{j=1}^{|z_{\caly'_k}|} 2^{l_{k,j}}}{|z_{\caly'_k}|!}  = \frac{\displaystyle \prod_{j=1}^{|z_{\caly'_k}|} 2^{l_{k,j}}}{|z_{\caly'_k}|!}
%     \leq \frac{\displaystyle \prod_{j=1}^{|z_{\caly'_k}|} 2^{\frac{k}{L'}\log \bl}}{|z_{\caly'_k}|!} = \frac{(2^{\frac{k}{L'}\log \bl})^{|z_{\caly'_k}|}}{|z_{\caly'_k}|!}. 
% \end{align*}
% %%%Thus, using known bounds \cite[Chapter 1.2.5, Exercise 24]{Knuth_factorialbound} on the factorial, we have 
% \begin{equation}
%     \frac{|\caly'_k|^{|\caly'_k|}}{e^{|\caly'_k|-1}} \leq |\caly'_k|! \leq \frac{|\caly'_k|^{^{|\caly'_k|}+1}}{e^{|\caly'_k|-1}}
% \label{eq:factorialboundonsizeofYk'} 
% \end{equation}
Using \eqref{eq:boundonsizeofyk'}, known bounds on the factorial ($a!\geq \frac{a^a}{e^a-1}$, see \cite[Chapter 1.2.5, Exercise 24]{Knuth_factorialbound}) and the fact that $|z_{\caly'_k}| = |\caly'_k|$, i.e., the number of erasure patterns in $z_{\caly'_k}$ is exactly corresponding to the number of reads in $\caly'_k$, we have 
%%%%%
\recalc{
\begin{align}
    &\frac{\displaystyle \prod_{j=1}^{|z_{\caly'_k}|} 2^{l_{k,j}}}{|z_{\caly'_k}|!}\nonumber \\
    & \leq \frac{2^{\sum_{j=1}^{|z_{\caly'_k}|}l_{k,j}}}{{(\frac{\bl}{\ExLenIsland}(q_{k,\bl}\ekn - \epsilon))^{\frac{\bl}{\ExLenIsland}(q_{k,\bl}\ekn - \epsilon)}}/{e^{(\frac{\bl}{\ExLenIsland}(q_{k,\bl}\ekn - \epsilon)) - 1}}} 
    % \\
 % &\leq \frac{1}{e}\cdot\frac{(2^{\frac{k}{L'}\log \bl})^{\frac{\bl}{\ExLenIsland}( q_{k,\bl}\ekn + \epsilon)}e^{\frac{\bl}{\ExLenIsland}(q_{k,\bl}\ekn - \epsilon)}}{(\frac{\bl}{\ExLenIsland}(q_{k,\bl}\ekn - \epsilon))^{\frac{\bl}{\ExLenIsland}(q_{k,\bl}\ekn - \epsilon)}} \\
 % &= \frac{((2^{\frac{k}{L'}\log \bl})\frac{\bl}{\ExLenIsland}(q_{k,\bl}\ekn - \epsilon))^{\frac{\bl}{\ExLenIsland}\epsilon}}{e^{1+\frac{\bl}{\ExLenIsland}\epsilon}}\cdot\frac{(2^{\frac{k}{L'}\log \bl}e)^{\frac{\bl}{\ExLenIsland}( q_{k,\bl}\ekn)}}{(\frac{\bl}{\ExLenIsland}(q_{k,\bl}\ekn - \epsilon))^{\frac{\bl}{\ExLenIsland}(q_{k,\bl}\ekn)}}
 \triangleq \sigma_{z_{\caly'_k}}.
 \label{eqref:sigma_recalc}
 \end{align}
%%%%
Thus, from \eqref{eqref:sigma_recalc}, we get
\begin{align}
    H&(\caly'_k|Z_{\caly'_k} = z_{\caly'_k},\bar{\cal E}_{k,\bl})\nonumber \\
    &\leq \log(\sigma_{z_{\caly'_k}} \displaystyle \prod_{j=1}^{|z_{\caly'_k}|} 2^{- \rho_{k,j}}) 
\leq \log(\sigma_{z_{\caly'_k}}) - \sum\limits_{j = 1}^{|z_{\caly'_k}|}\rho_{k,j}.
\label{eqn:boundforspecificerasurepattern}
\end{align}
}

\recalc{
Using \eqref{eqn:boundforspecificerasurepattern} in \eqref{eqn:expansionofyk'givenZyk'andScriptEtilde}, we have
%%%
\begin{align}   
\nonumber H&(\caly'_k|Z_{\caly'_k},\bar{\cal E}_{k,\bl}) \\
\nonumber
&= \sum_{z_{\caly'_k}}\Pr(Z_{\caly'_k} = z_{\caly'_k} | \bar{\cal E}_{k,\bl})H(\caly'_k|Z_{\caly'_k} = z_{\caly'_k},\bar{\cal E}_{k,\bl})\\
\nonumber
 &\leq \sum_{z_{\caly'_k}}\Pr(Z_{\caly'_k} = z_{\caly'_k} | \bar{\cal E}_{k,\bl})\left(\log(\sigma_{z_{\caly'_k}}) - \sum\limits_{j = 1}^{|z_{\caly'_k}|}\rho_{k,j}\right)\\
 % \\   \nonumber
 % &=\sum_{z_{\caly'_k}}\Pr(Z_{\caly'_k} = z_{\caly'_k} | \bar{\cal E}_{k,\bl})\log(\sigma) - \sum_{z_{\caly'_k}}\Pr(Z_{\caly'_k} = z_{\caly'_k} | \bar{\cal E}_{k,\bl})\sum\limits_{j = 1}^{|z_{\caly'_k}|}\rho_{k,j}
  &= \expect\left[\sum_{\substack{i=1}}^{|\caly'|}N_i~|~~\bar{\cal E}_{k,\bl}\right]\nonumber \\
  &+\frac{\bl\epsilon}{\ExLenIsland}\left(\log\left(\frac{\bl}{\ExLenIsland}(q_{k,\bl}\ekn - \epsilon)\right)-\log e\right) \nonumber \\
  &~~~~~~~+\frac{\bl}{\ExLenIsland}q_{k,\bl}\ekn\left(\log e - \log\left(\frac{\bl}{\ExLenIsland}( q_{k,\bl}\ekn - \epsilon)\right)\right) \nonumber \\
  &~~~~~~~~~- \log e-\sum\limits_{j = 1}^{|z_{\caly'_k}|}\Pr(Z_{\caly'_k} = z_{\caly'_k} | \bar{\cal E}_{k,\bl})\sum\limits_{j = 1}^{|z_{\caly'_k}|}\rho_{k,j}.
\label{25-recalc}
\end{align} 
}

Now, we  bound $\sum_{z_{\caly'_k}}\Pr(Z_{\caly'_k} = z_{\caly'_k} | \bar{\cal E}_{k,\bl})\sum\limits_{j = 1}^{|z_{\caly'_k}|}\rho_{k,j}$. Denoting the total number of erasures in $z_{\caly'_k}$ as $\rho(z_{\caly'_k})$, i.e., $\rho(z_{\caly'_k}) \triangleq \sum\limits_{j = 1}^{|z_{\caly'_k}|}\rho_{k,j},$
we have 

\recalc{
\begin{align}
\nonumber
    \sum_{z_{\caly'_k}}&\Pr(Z_{\caly'_k} = z_{\caly'_k} | \bar{\cal E}_{k,\bl})\sum\limits_{j = 1}^{|z_{\caly'_k}|}\rho_{k,j} \\
    \nonumber &= \sum_{z_{\caly'_k}}\Pr(Z_{\caly'_k} = z_{\caly'_k} | \bar{\cal E}_{k,\bl})\rho(z_{\caly'_k}) \\
    \nonumber
    &= \sum\limits_{\rho = 0}^{\bl}\rho \sum_{z_{\caly'_k} : \rho(z_{\caly'_k}) = \rho}\Pr(Z_{\caly'_k} = z_{\caly'_k}| \bar{\cal E}_{k,\bl}) \\
    \nonumber
    &=  \sum\limits_{\rho = 0}^{\bl}\rho \Pr(\rho(Z_{\caly'_k}) = \rho| \bar{\cal E}_{k,\bl})
    \\
    \nonumber
    &={\expect}(\rho(Z_{\caly'_k})|\bar{\cal E}_{k,\bl})\\
    % &={\expect}\left(\delta\cdot\sum_{j\in |z_{\caly'_k}|}l_{k,j}~|~\bar{\cal E}_{k,\bl}\right)\\
    &\stackrel{(a)}{=}\expect\left[\sum_{\substack{i=1}}^{|\caly'|}N_i~|~~\bar{\cal E}_{k,\bl}\right]\delta_e
    \label{eqn:30-recalc}
\end{align}
where in $(a)$, we use the linearity of expectation applied to the indicators of the erasures at each location. Using \eqref{eqn:30-recalc} in \eqref{25-recalc},
% \pk{include using which equations in what we get the below}
\begin{align}
H&(\caly'_k|Z_{\caly'_k},\bar{\cal E}_{k,\bl}) \\
&\leq (1-\delta_e)\expect\left[\sum_{\substack{i=1}}^{|\caly'|}N_i~|~~\bar{\cal E}_{k,\bl}\right]\nonumber 
\\&~~~~~~~~~~~~~~~+\frac{\bl\epsilon}{\ExLenIsland}\left(\log\left(\frac{\bl}{\ExLenIsland}(q_{k,\bl}\ekn - \epsilon)\right)-\log e\right) \nonumber \\
  &~~~~~~~~~~+\frac{\bl}{\ExLenIsland}q_{k,\bl}\ekn\left(\log e - \log\left(\frac{\bl}{\ExLenIsland}( q_{k,\bl}\ekn - \epsilon)\right)\right) - \log e.
\label{eq:boundoncaly'kgivenzcaly'kANDscriptE-recalc}
\end{align}
}

We now bound the second term in the R.H.S of \eqref{eq:11}, $H(Z_{\caly'_k}|\bar{\cal E}_{k,\bl})$, from above. We can write,
\begin{align}
    H(Z_{\caly'_k}|\bar{\cal E}_{k,\bl}) &= I(Z_{\caly'_k};\vX,\Gamma^\bl|\bar{\cal E}_{k,\bl}) + H(Z_{\caly'_k}|\bar{\cal E}_{k,\bl},\vX,\Gamma^\bl).
    \label{eq:29}
\end{align}
Observe that $I(Z_{\caly'_k};\vX,\Gamma^\bl|\bar{\cal E}_{k,\bl})$ represents the information about the lengths of the patterns in $Z_{\caly'_k}$ and the cardinality $|Z_{\caly'_k}|$, given $\bar{\cal E}_{k,\bl}$, as is this information is what can be gathered from $\vX$ and $\Gamma^\bl$. Note that, given $\bar{\cal E}_{k,\bl}$, the cardinality $|Z_{\caly'_k}|=|\caly'_k| \in (\frac{\bl}{\ExLenIsland} ( q_{k,\bl}\ekn -\epsilon),\frac{\bl}{\ExLenIsland} ( q_{k,\bl}\ekn + \epsilon))$, and thus $|Z_{\caly'_k}|$ can take one of at most $\frac{2n\epsilon}{\ExLenIsland}$ possible values. Further, there are at most $\frac{\log \bl}{L'}$ possible values for the lengths of the error patterns in $Z_{\caly'_k}$, by definition of $\caly'_k$. Thus,  we have

\begin{align}
    I&(Z_{\caly'_k};\vX,\Gamma^\bl|\bar{\cal E}_{k,\bl})\nonumber\\ &\leq \left(\log \frac{\log \bl}{L'}\right)\frac{\bl}{\ExLenIsland} ( q_{k,\bl}\ekn + \epsilon) + \log \left(\frac{2n\epsilon}{\ExLenIsland}\right),
    \label{eq:30}
\end{align}
where the first term in the R. H. S. is product of a bound on the entropy in length of each read in $Z_{\caly'_k}$ and the upper bound on $|\caly'_k|$ when $\bar{\cal E}_{k,\bl}$ holds, while the second term bounds the random cardinality $|Z_{\caly'_k}|$.

% \pk{March 6: We can show that as $\bl\to \infty$ and $\epsilon\to 0$, $\sum_{k=1}^\infty I(z_{\caly'_k};X^\bl,T^\bl|\bar{\cal E}_{k,\bl})\to 0$. [Continue here if you wish]
% \begin{align*}
% I(z_{\caly'_k};X^\bl,T^\bl|\bar{\cal E}_{k,\bl})
% \end{align*}
% }

% \pk{March 6: This additional term  $I(z_{\caly'_k};X^\bl,T^\bl|\bar{\cal E}_{k,\bl})$ has to get reflected in the subsequent equations.}

% Now we have to find $H(z_{\caly'_k}|\bar{\cal E}_{k,\bl},X^\bl,T^\bl)$,

% \begin{align}
%     H(z_{\caly'_k}|\bar{\cal E}_{k,\bl},X^\bl,T^\bl) &\leq H_2(\delta)|\caly'_k|\frac{k}{L'}\log \bl \leq H_2(\delta)\frac{\bl}{\ExLenIsland}( q_{k,\bl}\ekn + \epsilon) \frac{k}{L'}\log \bl 
%     \label{eq:31}
% \end{align}

Using \eqref{eq:30} in \eqref{eq:29} we get,

\begin{align}
     H(Z_{\caly'_k}|\bar{\cal E}_{k,\bl}) &\leq \left(\log \frac{\log \bl}{L'}\right)\frac{\bl}{\ExLenIsland} ( q_{k,\bl}\ekn + \epsilon) \nonumber \\
     &~~~~+ \log\left(\frac{2n\epsilon}{\ExLenIsland}\right) + H(Z_{\caly'_k}|\vX,\Gamma^\bl,\bar{\cal E}_{k,\bl})
     %\\
     %&\leq H(Z_{\caly'_k}|\vX,\Gamma^\bl,\bar{\cal E}_{k,\bl})+o(\bl),
     % \label{eq:32}
\end{align}
% Note that for $\epsilon=\log\bl\left(\sqrt{\frac{\bl}{2\ExLenIsland}}\right)^{-1}$ (following \eqref{eqn:minepsiloncondition}),  and 
 Since $\ExLenIsland$ is at least $\normrl \log n$, we have 
\begin{align}
     \frac{1}{n}&H(Z_{\caly'_k}|\bar{\cal E}_{k,\bl}) \nonumber\\&\leq \frac{1}{n}H(Z_{\caly'_k}|\vX,\Gamma^\bl,\bar{\cal E}_{k,\bl})+\tilde{f}_1(\bl)q_{k,\bl}\ekn+\tilde{f}_2(\bl)\epsilon,
     %\\
     %&\leq H(Z_{\caly'_k}|\vX,\Gamma^\bl,\bar{\cal E}_{k,\bl})+o(\bl),
     \label{eq:32}
\end{align}
where both $\tilde{f}_1(\bl)$ and $\tilde{f}_2(\bl)$ are $O(\log(\log\bl)/\log\bl).$ 
%\pk{Even if we assume $L'=\log J=\log\log n$ here, this will be true. And hopefully this specific ineq doesn't create an issue even when summing over $k=1..J=\log n$ , multiplying by $\epsilon$ (equal to $\log n/\sqrt{n/\ExLenIsland}$) and dividing by $n$}

\recalc{
Therefore, using \eqref{eq:boundoncaly'kgivenzcaly'kANDscriptE-recalc}, \eqref{eq:32} and \eqref{eq:11}, we have
\begin{align}
\nonumber 
&\frac{1}{\bl}H(\caly'_k|\bar{\cal E}_{k,\bl})\\
&\leq \frac{1}{\bl}(1-\delta_e)\expect\left[\sum_{\substack{i=1}}^{|\caly'|}N_i~|~~\bar{\cal E}_{k,\bl}\right]\nonumber \\
&+\frac{\epsilon}{\ExLenIsland}\left(\log\left(\frac{\bl}{\ExLenIsland}(q_{k,\bl}\ekn - \epsilon)\right)-\log e\right) \nonumber \\
  &+\frac{1}{\ExLenIsland}q_{k,\bl}\ekn\left(\log e - \log\left(\frac{\bl}{\ExLenIsland}( q_{k,\bl}\ekn - \epsilon)\right)\right) - \frac{\log e}{\bl}\nonumber\\
  &+ \frac{1}{\bl}H(Z_{\caly'_k}|\vX,\Gamma^\bl,\bar{\cal E}_{k,\bl})+\tilde{f}_1(n)q_{k,\bl}\ekn+\tilde{f}_2(n)\epsilon. \label{35-recalc}
\end{align}
}

\recalc{
Let $M_k \triangleq \frac{q_{k,\bl}\ekn}{\ExLenIsland}$. 
Let \begin{align}
    \label{eqn:defnofQ-recalc}
    Q_{k,n,\epsilon} &\triangleq \frac{1}{\ExLenIsland}\left(\log\left(\frac{\bl}{\ExLenIsland}(q_{k,\bl}\ekn - \epsilon)\right)-\log e\right).
%%%%
\end{align} 

Using these in \eqref{35-recalc}, we get 

\begin{align}
    \frac{1}{\bl}&H(\caly'_k|\bar{\cal E}_{k,\bl})\nonumber \\&\leq \frac{1}{\bl}(1-\delta_e)\expect\left[\sum_{\substack{i=1}}^{|\caly'|}N_i~|~~\bar{\cal E}_{k,\bl}\right]\nonumber\\
    &~~~+M_k\ekn\left(\log e - \log\left(\bl M_k - \frac{\bl}{\ExLenIsland}\epsilon)\right)\right)\nonumber\\
  &~~~+ \frac{1}{\bl}H(Z_{\caly'_k}|\vX,\Gamma^\bl,\bar{\cal E}_{k,\bl})\nonumber\\
  &~~~~~+\epsilon (\tilde{f}_2(\bl)+Q_{k,n,\epsilon})+\tilde{f}_1(\bl)q_{k,\bl}\ekn-\frac{\log e}{\bl}. \label{eqn:38-recalc}
\end{align}
%%%%
% \pk{Apr18:Till this point no assumption on J. Assumptions on J follow in calcs below.}
% 
}

\recalc{
We now show that, for any fixed $J\leq \beta\log\bl$ (for any constant $\beta$) and $L'$, with $\epsilon=\log\bl\left(\sqrt{\frac{\bl}{2\ExLenIsland}}\right)^{-1}$ (chosen based on \eqref{eqn:probofcalEgoesto0}), 
%%%%
\begin{align}
\label{eqn:unnecessarypartsgotozero}
\lim_{n\to \infty}\sum_{k=1}^{J-1}\left(\epsilon (\tilde{f}_2(\bl)+Q_{k,n,\epsilon})+\tilde{f}_1(\bl)q_{k,\bl}\ekn-\frac{\log e}{\bl}\right) = 0.
%~\text{as}~\bl\to\infty.
\end{align}
%%%%
By definition $\tilde{f}_1(\bl)$ and $\tilde{f}_2(\bl)$ are $O(\log\log\bl/\log\bl)$. Using these and \eqref{eq:sumofprobabilitiesofIslands}, under our choices for $L'$ and $J$, it can be seen that 
\begin{align}
\label{eqn:oneofunnecessaryparts}
    \sum_{k=1}^{J-1}\left(\epsilon \tilde{f}_2(\bl)+\tilde{f}_1(\bl)q_{k,\bl}\ekn-\frac{\log e}{n}\right)=O(J(\log\bl\log\log\bl)^2/\sqrt{\bl}).
\end{align}

Now we consider the term $\sum_{k=1}^{J-1}\epsilon Q_{k,n,\epsilon}$.
\begin{align}
    \sum_{k=1}^{J-1}\epsilon Q_{k,n,\epsilon}&=\frac{\epsilon}{\ExLenIsland}\sum_{k=1}^{J-1}\left(\log\left(\frac{\bl}{\ExLenIsland}(q_{k,\bl}\ekn - \epsilon)\right)-\log e\right) \nonumber\\
    \label{eqn:unneccessarypart2-recalc}
    &\leq \epsilon\log\bl\sum_{k=1}^{J-1}\left(\log\bl-\log e\right) =O(J\log^3\bl/\sqrt{\bl}).
\end{align}
}

\recalc{
Thus, from \eqref{eqn:oneofunnecessaryparts} and \eqref{eqn:unneccessarypart2-recalc}, we have \eqref{eqn:unnecessarypartsgotozero}. Using \eqref{eqn:unnecessarypartsgotozero} and \eqref{eqn:38-recalc}, we have

\begin{align}
    \sum_{k=1}^{J-1}&\frac{1}{\bl}H(\caly'_k|\bar{\cal E}_{k,\bl})\nonumber\\
    &\leq \sum_{k=1}^{J-1}\frac{1}{\bl}(1-\delta_e)\expect\left[\sum_{\substack{i=1}}^{|\caly'|}N_i~|~~\bar{\cal E}_{k,\bl}\right]\nonumber\\
&+\sum_{k=1}^{J-1}\frac{q_{k,\bl}\ekn}{\ExLenIsland}\left(\log e - \log\left(\bl M_k - \frac{\bl}{\ExLenIsland}\epsilon)\right)\right)\nonumber\\
  &~~~~~~~~~~~~~~~~+ \sum_{k=1}^{J-1}\frac{1}{\bl}H(Z_{\caly'_k}|\vX,\Gamma^\bl,\bar{\cal E}_{k,\bl})+o(1)\nonumber\\
  &=\frac{1}{\bl}(1-\delta_e)\sum_{k=1}^{J-1}\expect\left[\sum_{\substack{i=1}}^{|\caly'|}N_i~|~~\bar{\cal E}_{k,\bl}\right]\nonumber\\
  &+\sum_{k=1}^{J-1}\frac{q_{k,\bl}\ekn}{\ExLenIsland}\left(\log e -\log\bl- \log\left(\frac{q_{k,n}\ekn -\epsilon}{\ExLenIsland}\right)\right)\nonumber\\
  \label{eqn:almostthere}&~~~+ \frac{1}{\bl}\sum_{k=1}^{J-1}H(Z_{\caly'_k}|\vX,\Gamma^\bl,\bar{\cal E}_{k,\bl})+o(1).
\end{align}
}
Combining \eqref{eqn:almostthere} with \eqref{eqn:initialboundonEntropyofcalyk'} completes the proof of the claim. 
%%%%
  \section{Proof of Claim \ref{claim:final}}
\label{app:proofoffinalclaim}
Observe that we can write 
\begin{align*}
\expect\left[\sum_{\substack{i=1}}^{|{\caly'}|}N_i\right]=&\expect_{\indicatorRV_{{\cal E}_{k,\bl}}}\left[\expect\left[\sum_{\substack{i=1}}^{|{\caly'}|}N_i~|~\indicatorRV_{{\cal E}_{k,\bl}}\right]\right]\\
&=\expect\left[\sum_{\substack{i=1}}^{|{\caly'}|}N_i~|~\bar{\cal E}_{k,\bl}\right]\Pr(\bar{\cal E}_{k,\bl})\\
&~~+\expect\left[\sum_{\substack{i=1}}^{|{\caly'}|}N_i~|~{\cal E}_{k,\bl}\right]\Pr({\cal E}_{k,\bl}).
\end{align*}
Now, using \eqref{eqn:probofcalEgoesto0}, we see that 
%%%
\begin{align}
\label{eqn:coverageunderconditioning}
    \lim_{\bl\to\infty}\frac{1}{\bl}\expect\left[\sum_{\substack{i=1}}^{|{\caly'}|}N_i\right]=\lim_{\bl\to\infty}\frac{1}{\bl}\expect\left[\sum_{\substack{i=1}}^{|{\caly'}|}N_i~|~\bar{\cal E}_{k,\bl}\right].
\end{align}    
%%%

\recalc{
Using \eqref{eqn:coverageunderconditioning} and Claim \ref{claim:bounonHcalyk'byn}, we thus have

\begin{align*}
    &\lim_{\bl\to \infty}\left(\sum_{k=1}^{J-1}\frac{H(\caly'_k)}{\bl}-\sum_{k=1}^{J-1} \frac{H(Z_{\caly'_k}|\vX,\Gamma^\bl,\bar{\cal E}_{k,\bl})}{\bl} \right)\nonumber\\ 
    &= \lim_{\bl \to \infty} \sum_{k=1}^{J-1}\frac{1}{\bl}(1-\delta_e)\expect\left[\sum_{\substack{i=1}}^{|\caly'|}N_i~|~~\bar{\cal E}_{k,\bl}\right]\nonumber\\
    &~~~~~~~- \lim_{\bl\to \infty}\frac{\log \bl}{\ExLenIsland}\sum_{k=1}^{J-1}q_{k,\bl}\ekn \nonumber\\ 
    &\leq \left(1-\delta_e\right)\lim_{\bl\to\infty}\sum_{k=1}^{J-1}\frac{1}{\bl}\expect\left[\sum_{\substack{i=1}}^{|{\caly'}|}N_i\right] - \lim_{\bl\to \infty}\frac{\log \bl}{\ExLenIsland}\sum_{k=1}^{J-1}q_{k,\bl}\ekn \nonumber\\
    &\leq \left(1-\delta_e\right)\lim_{\bl\to\infty}\sum_{k=1}^{\infty}\frac{1}{\bl}\expect\left[\sum_{\substack{i=1}}^{|{\caly'}|}N_i\right] - \lim_{\bl\to \infty}\frac{\log \bl}{\ExLenIsland}\sum_{k=1}^{J-1}q_{k,\bl}\ekn \nonumber\\
    &= \left(1-\delta_e\right)\lim_{\bl\to\infty}\frac{1}{\bl}\expect\left[\sum_{\substack{i=1}}^{|{\caly'}|}N_i\right] - \lim_{\bl\to \infty}\frac{\log \bl}{\ExLenIsland}\sum_{k=1}^{J-1}q_{k,\bl}\ekn \nonumber\\
    &\stackrel{(a)}{\leq} \left(1-\delta_e\right)\lim_{\bl\to\infty}\frac{1}{\bl}\expect\left[\sum_{\substack{i=1}}^{|{\caly'}|}N_i\right] - \lim_{\bl\to \infty}\frac{\log\bl }{\bl}\expect[\noofreads']\sum_{k=1}^{J-1}q_{k,\bl}\ekn,
\end{align*}
where $(a)$ holds as $\expect[\noofreads']\cdot\Lambda=\expect_{K'}[\sum_{i=1}^{K'}\expect[N_i]|K']\leq \bl$. This proves the claim. 

}

\section{Proof of Claim \ref{claim:entropyofy'GreaterthanJgoesto0}}
\label{app:proof:claim:entropyofy'GreaterthanJgoesto0}
% \pk{JULY 2025, POST ISIT: NOTE: Parts of this proof were changed a bit from the conference-submitted version to the camera-ready full-version. Essentially, wherever $\sum_{i=1}^{|{\caly'}|}N_i$ appears, in all those places $Y_i'\in \caly' $ was essentially there in the summation, now it seemed irrelevant and was removed in THIS FILE.}
Consider the quantity $\frac{1}{\bl} \sum_{i=1}^{|\caly'|} N_i \indicatorRV_{\{N_i \geq \gamma \log \bl\}}$. This essentially measures the `coverage' (the number of bits in $\vX$ that were covered, though not necessarily visibly so) by the islands of length at least $\gamma \log \bl$. For some $\gamma,\zeta>0$, define the event 
% \pk{use $\zeta$ in the place of $\delta$ below}
\begin{align*}
    T_{\gamma,\zeta} \triangleq \Big\{\frac{1}{\bl} \sum_{i=1}^{|\caly'|} N_i \indicatorRV_{\{N_i \geq \gamma \log \bl\}}> {\expect}\left[\frac{N_i}{\ExLenIsland} \indicatorRV_{\{N_i \geq \gamma \log \bl\}}\right] + \zeta \Big\}. 
    \end{align*}

%  From Lemma 6 in the \href{https://arxiv.org/abs/2407.05544}{paper} \cite{}, we see that $\lim_{\bl\to\infty}\Pr(T_{\gamma,\zeta}) \to 0$,\pk{Have to verify if this is exact reference and we don't need to reprove.}
    % \pk{March 3: We can then show by standard arguments?
    % \begin{align*}
    %     \lim_{\bl\to \infty} Pr(T_{\gamma,\zeta}) & \to 0.
    % \end{align*}
    % }
    
From \cite[Lemma 1 and proof of Lemma 4]{ravi_recovering_message_incomplete_set}, the following concentration result is obtained. 
\begin{align}
\lim_{n\to\infty}\Pr(T_{\gamma,\zeta})=0.
\end{align}

Observe that for $\gamma=\frac{J-1}{L'}$, the event $\{N_i \geq \gamma \log \bl\}=\{Y_i'\in \caly'_{\geq J}\},$ where $Y_i'$ is the $i^{\tth}$ island in $\caly'$.
We recall that $H(\caly'_{\geq J}|\bar{T}_{\gamma,\zeta})\leq H(\caly'|\vX,\Gamma^\bl,Z_{\caly'},\bar{T}_{\gamma,\zeta})+H(\vX,\Gamma^\bl,Z_{\caly'}|\bar{T}_{\gamma,\zeta})=0+H(\vX,\Gamma^\bl,Z_{\caly'}|\bar{T}_{\gamma,\zeta})\leq 3n\left({\expect}\left(\frac{N_i\indicatorRV_{\{Y_i'\in \caly'_{\geq J}\}}}{\ExLenIsland}\right)+\zeta\right)$. By similar arguments, we can show that $H(\caly'_{\geq J}|T_{\gamma,\zeta})\leq 3n.$  Using these observations, and bounding $H(\caly'_{\geq J})$ by the joint entropy $H(\caly'_{\geq J},\indicatorRV_{T_{\gamma,\zeta}})$, we get the following.
% \pk{remove unnecessary numbering from the eqns below using nonumber command}
% \pk{Replace expectation $E$ everywhere by ${\expect}$}
\begin{align*}
\nonumber
    H(\caly'_{\geq J}) &\leq H(\caly'_{\geq J} | \bar{T}_{\gamma,\zeta}) + H(\caly'_{\geq J} | T_{\gamma,\zeta}) \Pr(T_{\gamma,\zeta}) + 1   \\
    % &\stackrel{(a)}{\leq} H(\caly'_{\geq J} | \bar{T}_{\gamma,\zeta})  + 1 \\
    &\stackrel{}{\leq} 3n\left({\expect}\left(\frac{N_i\indicatorRV_{\{Y_i'\in \caly'_{\geq J}\}}}{\ExLenIsland}\right)+\zeta\right) + 3n \Pr(T_{\gamma,\zeta})+1 \\
&\stackrel{(a)}\leq3n\left(\sqrt{{\expect}(N_1^2)\Pr(Y_i'\in \caly'_{\geq J})}/\ExLenIsland + \zeta\right) +o(\bl) \\
    &\leq3n\left(\sqrt{{\expect}(N_1^2)\Pr\left(N_i\geq \frac{J-1}{L'}\log \bl\right)}/\ExLenIsland + \zeta\right) +o(\bl) \\
    &\stackrel{(b)}\leq 3n\left(\frac{\sqrt{{\expect}(N_1^2)\ExLenIsland L'}}{\ExLenIsland\sqrt{(J-1)\log \bl}} + \zeta\right) +o(\bl) \\
    &\stackrel{(c)}\leq3n(S\sqrt{L'/(J-1)} + \zeta) + o(\bl).
\end{align*}
%%%
% \pk{Write reasoning for 3rd inequality}
  $(a)$ is due to Cauchy-Schwartz Inequality, $(b)$ is due to Markov's Inequality, $(c)$ is true for $S\triangleq \sqrt{\frac{{\expect}(N_1^2)}{\ExLenIsland \log \bl}}$. Further, this $S$ is finite, since $(i)$ $\lim_{\bl\to\infty}\frac{\log \bl}{\expect[N_1]} \in (0,\infty)$ and (ii) $\expect[N_1^2]/(\log \bl)^2$ is finite and bounded, as observed in the beginning of Subsection \ref{subsec:partitionandproperties}. Dividing by $\bl$ on both sides, we have proved the statement. 

%%%
\end{document}